\documentclass[useAMS,usenatbib]{mn2e}
\usepackage{epsfig,amsmath,amssymb,amsfonts,mathrsfs,latexsym,graphicx}

\usepackage{hyperref}
\usepackage{natbib}

\usepackage{longtable}

\usepackage{color}

\include{journals}
\graphicspath{{images/}}

\newcommand{\eg}{e.g.\ }

\newcommand{\Msun}{{\rm M}_{\odot}}

\newcommand{\kms}{km\,s$^{-1}$}
\newcommand{\ergs}{erg s$^{-1}$}

\newcommand{\CII}{C~{\sc ii}}
\newcommand{\CIII}{C~{\sc iii}}
\newcommand{\OI}{O~{\sc i}}

\newcommand{\NaI}{Na~{\sc i}}

\newcommand{\MgII}{Mg~{\sc ii}}

\newcommand{\SII}{S~{\sc ii}}
\newcommand{\SIII}{S~{\sc iii}}

\newcommand{\SiII}{Si~{\sc ii}}
\newcommand{\SiIII}{Si~{\sc iii}}

\newcommand{\CaII}{Ca~{\sc ii}}

\newcommand{\FeII}{Fe~{\sc ii}}
\newcommand{\FeIII}{Fe~{\sc iii}}

\newcommand{\CoII}{Co~{\sc ii}}
\newcommand{\CoIII}{Co~{\sc iii}}

\newcommand{\Feff}{$^{54}$Fe}
\newcommand{\Fefs}{$^{56}$Fe}
\newcommand{\Cofs}{$^{56}$Co}
\newcommand{\Nifs}{$^{56}$Ni}
\newcommand{\Nife}{$^{58}$Ni}

\newcommand{\Deltam}{$\Delta m_{15}(B)$}

\def\gsim{\mathrel{\rlap{\lower 4pt \hbox{\hskip 1pt $\sim$}}\raise 1pt \hbox {$>$}}}
\def\lsim{\mathrel{\rlap{\lower 4pt \hbox{\hskip 1pt $\sim$}}\raise 1pt \hbox {$<$}}}

\title[Abundance stratification in SN\,1991T]{Abundance stratification in 
Type Ia Supernovae - IV: the luminous, peculiar SN\,1991T }

\author[M. Sasdelli et al.]{Michele Sasdelli$^{1}$\thanks{E-mail: 
sasdelli@mpa-garching.mpg.de}, P. A. Mazzali$^{2,1,3}$, E. Pian$^{4,5}$,
K. Nomoto$^{6,7}$, S. Hachinger$^8$,
\newauthor E. Cappellaro$^3$, S. Benetti$^3$ 
\\
$^1$Max-Planck Institut f\"ur Astrophysik, Karl-Schwarzschildstr. 1, D-85748 
Garching, Germany \\
$^2$Astrophysics Research Institute, Liverpool John Moores University,
Liverpool L3 5RF, UK \\
$^3$INAF-Osservatorio Astronomico, vicolo dell'Osservatorio, 5, I-35122 
Padova, Italy \\
$^4$INAF-IASF-Bo, via Gobetti, 101, I-40129 Bologna, Italy \\
$^5$Scuola Normale Superiore, Piazza dei Cavalieri 7, 56126 Pisa, Italy \\
$^6$Kavli Institute for the Physics and Mathematics of the Universe (WPI), 
The University of Tokyo, Kashiwanoha 5-1-5, Kashiwa, \\
Chiba 277-8583, Japan \\
$^7$Hamamatsu Professor\\
$^8$Institut f\"ur Theoretische Physik und Astrophysik, 
Universit\"at W\"urzburg, Emil-Fischer-Str. 31, 97074 W\"urzburg, Germany\\
}

\begin{document}

\date{Accepted ... Received ...; in original form ...}

\pagerange{\pageref{firstpage}--\pageref{lastpage}} \pubyear{2014}

\maketitle

\label{firstpage}

\begin{abstract}

The abundance distribution of the elements in the ejecta of the peculiar,
luminous Type Ia supernova (SN\,Ia) 1991T is obtained modelling spectra 
{from before maximum light until a year after the explosion, with the method of}
``Abundance Tomography".
SN 1991T is different from other slowly declining SNe\,Ia (e.g. SN\,1999ee)
in having a weaker \SiII\,6355 line and strong features of
iron group elements before maximum. The distance to the SN is investigated along
with the abundances and the density profile.
The ionization transition that happens {around maximum sets a strict upper limit on the luminosity}.
 Both W7 and the WDD3 delayed detonation
model are tested. WDD3 is found to provide marginally
better fits. In this model the core of the ejecta is dominated by stable Fe with a mass of
about $0.15\Msun$, as in most SNe\,Ia. 
{The layer above is mainly \Nifs\ up to  $v\sim10000$\,\kms\ ($\approx0.78\Msun$).
 A significant amount of \Nifs\ ($\sim3$\,\%) is located in
the outer layers. A narrow layer between 10000\,\kms\ and $\sim12000$\,\kms\ is
dominated by intermediate mass elements (IME), $\sim0.18\Msun$. This is small for a SN\,Ia. 
The high luminosity and the consequently high ionization, and the high \Nifs\ abundance at high velocities 
explain the peculiar
early-time spectra of SN\,1991T. The outer part is mainly
of oxygen, $\sim0.3\Msun$. Carbon lines are never detected,
yielding an upper limit of $0.01\Msun$ for C.
The abundances obtained with the W7 density model are qualitatively similar to those of the WDD3 model.}
 Different elements
are stratified with moderate mixing, resembling a delayed detonation.

\end{abstract}


\section{Introduction}

Type Ia supernovae (SNe\,Ia) are among the most luminous stellar explosions.
They have been successfully used as distance indicators providing the first
direct evidence for the accelerated expansion of the Universe \citep{Riess98,
Perlmutter98}. A most remarkable feature of SNe\,Ia is the relation between
absolute magnitude and the shape of the light curve, which makes it possible to
calibrate them using a single parameter. The quantity traditionally used is 
\Deltam, defined as the difference in $B$-band magnitude between maximum and
15 days later \citep{1993ApJ...413L.105P}. 

SNe\,Ia can also be classified according to their spectral features. The aspect
of the spectrum varies with continuity together with \Deltam\ for most of
SNe\,Ia, representing a variation in temperature and therefore luminosity
\citep{Nugent95,Hachinger06,Hachinger08}. However, \Deltam\ is not sufficient to
account for all the variability of SN\,Ia spectra. In particular, at the bright
end of the \Deltam\ distribution some supernovae are spectroscopically different
from the bulk of normal SNe\,Ia. The first SN of this type was SN\,1991T
\citep{1992ApJ...384L..15F,1992AJ....103.1632P}, which gave its name to a
subclass of peculiar, overluminous SNe\,Ia.  Understanding the reasons for the
peculiarities among the most luminous SNe\,Ia is important for the reliability
of SN\,Ia cosmology, because these luminous SNe are favoured by Malmquist bias
at high redshift. 

The spectra of 1991T-like supernovae are characterised at early times by the
weakness or absence of \CaII\ and \SiII\ lines, and instead show very strong
 \FeIII\ lines up to about maximum. After maximum, the spectra become 
increasingly similar to those of normal SNe\,Ia. By about one week after maximum
their spectra are almost indistinguishable from those of normal SNe\,Ia with
comparable \Deltam. 

Luminous, peculiar SNe\,Ia are also characterised by a slowly declining light
curve. SN\,1991T has a \Deltam\,$=0.94\pm0.05$ \citep{1999AJ....118.1766P}. No
1991T-like SN have been found with a \Deltam\, larger than $\sim 1$\,mag.
However, bright SNe\,Ia are not well explained by the single-parameter
picture because there are spectroscopically normal bright SNe\,Ia with similarly slow
light curve decline \citep[\eg SN\,1999ee,][]{2002AJ....124..417H,Mazzali05}.

The actual luminosity of SN\,1991T has been a matter of debate, mostly because of
the uncertainty in the distance to its host galaxy, NGC\,4527. The 
measurements using different methods show significant dispersion.
The most recent measurements obtained using Cepheid variables, the Tully-Fisher
relation and surface brightness fluctuation method are summarized in Table
 \ref{dist_table}.

\begin{table}
\caption{Distance to NGC\,4527}
\begin{tabular}{cccc}
\label{dist_table}

Distance ($\mu$) & $H_0$ & method       & source \\
\hline
$30.76 \pm 0.20$ & 62   & Cepheid      &  \cite{2006ApJ...653..843S} \\
$30.56 \pm 0.08$ & 73   & Cepheid      &   \cite{2001ApJ...547L.103G} \\ 
$32.03 \pm 0.40$ & 70   & Tully-Fisher &   \cite{2007AandA...465...71T} \\
$30.64 \pm 0.35$ & --    & Tully-Fisher & \cite{2009AJ....138..323T} \\
$30.26 \pm 0.09$ & 76   & SBF          &  \cite{2001AandA...368..391R} \\
\hline
\end{tabular}
\end{table}

Owing to the peculiar nature of SN\,1991T, its intrinsic colour and, consequently, the amount
of reddening, have been a matter of debate. Galactic reddening is known to
be quite low, $E(B-V)=0.02$ \citep[][via NED]{ 1998ApJ...500..525S, 2011ApJ...737..103S}, but the total
reddening to SN\,1991T is not negligible. \cite{1999AJ....118.1766P} estimated
$E(B-V)=0.14\pm0.05$. More recently, \cite{2004MNRAS.349.1344A} found a higher
value, $E(B-V)=0.22\pm0.05$. An even higher reddening ($E(B-V)=0.3$) was
obtained from the equivalent width (EW) of the interstellar \NaI\,D line 
\citep{1992ApJ...387L..33R}. However, using EW(\NaI\,D)\,$= 1.37$\AA\ from
\cite{1992ApJ...387L..33R} together with the coefficients from
\cite{2003fthp.conf..200T} for the relation between reddening and EW(\NaI\,D)
gives a somewhat lower value ($E(B-V)= 0.2$).
{According to \cite{2013ApJ...779...38P} the EW(\NaI\,D)
 can be a quite unreliable indicator of reddening except in the case of a completely absent line.}

\begin{table}
\caption{The spectra of SN\,1991T}
\begin{tabular}{cccc}
\label{spectra_table}
Phase &   JD &  Obs.  &   Spectral Range  \\
 (days) &   (2448000+) &  &  (\AA)   \\ %
\hline
$-13.2$  &  362.5  &  ESO1.5m+B\&C  & $3400-8600 $ \\ 
$-12.0$  &  363.7  &  ESO2.2m+EF2   & $3300-8100 $ \\ 
$-11.0$  &  364.7  &  ESO2.2m+EF2   & $3300-8100 $ \\ 
$-10.2$  &  365.5  &  ESO1.5m+B\&C  & $3100-9600 $ \\ 
$ -9.0$  &  366.7  &   ESO3.6m+EF1  & $3500-10000$ \\ 
$ -8.1$  &  367.6  &   ESO3.6m+EF1  & $3200-10000$ \\ 
$ -7.0$  &  368.7  &   ESO3.6m+EF1  & $3500-10000$ \\ 
$ -4.2$  &  371.5  &   CTIO         & $2400-10400$ \\ 
$ -1.9$  &  373.8  &   IUE          & $2000-3400$ (UV)    \\
$ -0.4$  &  376.1  &   IUE          & $2000-3400$  (UV)  \\
$ +5.8$  &  381.5  &  CTIO          & $3100-9700 $ \\ 
$ +8.8$  &  384.5  &  CTIO          & $3100-9700 $ \\ 
$+13.8$  &  389.5  &  CTIO          & $2300-7600 $ \\ 
$+282$ &  657.7  &  ESO3.6m+EF1   & $3600-9500 $ \\ 

\end{tabular}
\end{table}

Despite the uncertainties in reddening and distance, most estimates indicate
that SN\,1991T is more luminous than the average SN\,Ia.
\cite{1992ApJ...384L..15F} stated that SN\,1991T is at least 0.6 mag more
luminous than normal SNe\,Ia. \cite{1992ApJ...387L..33R} estimated SN\,1991T to
be at least one magnitude more luminous than typical SNe\,Ia. Such a high
luminosity would require a large production of \Nifs\
in SN\,1991T. For this reason the supernova has sometimes been suggested to be
a super-Chandrasekhar explosion \citep{1999MNRAS.304...67F}. On the other hand,
\cite{2004MNRAS.349.1344A} stated that SN\,1991T fits well in the
Phillips-relation and that its peculiarities are only spectroscopic.
Finally, \cite{2001ApJ...547L.103G} state that the luminosity of SN1991T is 
indistinguishable from normal Type Ia SNe.

\cite*{Mazzali95} modelled a series of early time spectra of SN\,1991T
in order to determine its properties. They found that the lack of lines of
singly ionized species at early times could be explained if a large amount of
\Nifs\ was assumed to be located in the outer layers of the ejecta. According to
them the combined effect of low IME abundance and high luminosity and
consequently high ionization, which favours doubly ionized species, explains
the absence of \FeII, \SiII\ and \CaII\ lines and the presence of strong
\FeIII\ lines. \cite{1992AJ....103.1632P} and \cite{Mazzali95}
suggested that Si is present in a shell above 10000\,km\,s$^{-1}$.
{\cite{1992ApJ...397..304J} suggested that in the outer 
 layer of the ejecta (above 14500 \kms) 
\Nifs\ is present and it is more abundant than iron.}
Qualitatively these results are explained by delayed detonation models
\citep{1991A&A...245..114K,1991A&A...245L..25K,1992ApJ...393L..55Y}. To explain
the large production of \Nifs\ in the framework of these models, the burning
needs a brief deflagration phase with a quick transition to detonation
\citep{1996ApJ...472L..81H,2001NewA....6..307P,2009Natur.460..869K}.

Here we revisit the work of \cite{Mazzali95} with the technique of
Abundance Tomography \citep{Stehle05,Mazzali08} using a rich sample of spectra 
(Table~\ref{spectra_table}). This includes the late-time
nebular spectra, which offer insight into the innermost layers of the SN.

The observing material that we use here derives mainly from the ESO Key Program
 on SNe of the early '90s \citep{1991ESOC...37..725C}.
The spectra near maximum were presented for the first time in 
\cite{1992ApJ...387L..33R} whereas the late time nebular spectrum was shown
 in \cite{2001ApJ...549L.215C}.
The spectral sequence was complemented with observations obtained at the
 Lick Observatory \citep{1992ApJ...384L..15F} and at CTIO \citep{1992AJ....103.1632P}.
The photometric measurements were retrieved from \cite{1994ApJ...434L..19S},
 \cite{1998AJ....115..234L} and, the ESO program from \cite{2004MNRAS.349.1344A}.
We refer to the original papers for details on data acquisition and reduction.

The paper is organised as follows: Section~\ref{Method} describes briefly the
codes employed in our work and the modelling strategy used to fit the supernova
spectra. Section~\ref{Best reddening and distance} illustrates the modelling
procedure for the photospheric spectra. For the distance that yields the best
results, we show the effect of using different density profiles: the
``standard'' W7 density profile \citep{1984ApJ...286..644N} and the delayed
detonation model WDD3 \citep{Iwamoto99}. Section~\ref{Nebular model} describes
the modelling of the nebular spectrum. In Section~\ref{Abundances of elements}
we show the resulting abundance distributions.
The distribution of elemental abundances is the main result of our work.
 In Section~\ref{Bolometric_light_curve} we compare a synthetic bolometric light curve
obtained from our best-fitting model with the one inferred from the photometry.
Section~\ref{Check_the_consistency_of_the_model} is devoted to a check of the
consistence of the model with the expected kinetic energy yield and with the
Phillips relation. In Section~\ref{Comparison with explosion models} we compare
the results of the modelling with some explosion models for SNe\,Ia. Finally,
Section~\ref{Conclusions} concludes the paper.

\section{Method}
\label{Method}

{In this Section we describe the radiation transport codes used to model
 the spectra and the bolometric light curve of SN\,1991T.}

\subsection{Photospheric phase}

In the photospheric phase we use the Monte Carlo spectrum synthesis code
developed by \cite{Mazzali93} and later improved including photon
branching \citep{1999AA...345..211L,Mazzali2000}. In the first few weeks after
explosion, the optical depth of visible light reaches unity at a large radius,
including most of the ejected mass. The code approximates the emission of light
as black body emission coming from a photosphere. For
each spectrum in a time series, the photosphere recedes in velocity or
equivalently in mass coordinates. The photospheric approximation is reasonable if the
optical depth is similar at all wavelengths, and if the 
$\gamma$-ray deposition at radii larger than the photosphere is not too
important. These approximations are satisfied in the early phase and gradually
they begin to fail starting about two weeks after maximum.

The parameters of the model are, for each spectrum, the radius of the
photosphere, the luminosity of the supernova and the abundances of the elements
above the photosphere. We assume the density distribution of the W7 explosion
model or, alternatively, that of a delayed detonation model, and hence
density is not a free parameter.

The distance modulus and the reddening are used to scale the computed spectrum to the
observed flux. In previous papers
in this series the distance and reddening were kept fixed. Here, allowing for the
uncertainties in these quantities and the difficulties in fitting some spectral
features, we include these quantities among the free parameters. The effect of distance
and of reddening are similar in the resulting spectrum. We varied
both parameters probing the range of uncertainties of the measurements. In practice this
leads to a new, spectroscopic estimate of the luminosity of SN\,1991T.

\subsection{Nebular phase}
\label{Nebular_phase}

We model one of the available spectra of SN\,1991T using a non-local thermodynamic
equilibrium code similar to the description of \citet{Axelrod80}. The code
first computes the propagation and deposition of the $\gamma$-rays and positrons
produced in the decay of \Nifs\ to \Cofs\ and hence \Fefs, based on a density
and abundance profile, with a Monte Carlo scheme, as outlined for example in
\citet{Mazzali07}. The heating following this energy deposition is then
balanced by cooling via line emission. Both forbidden and low-lying permitted
transitions are included in our treatment. Line emission is
assumed to be homogeneous within each velocity shell in which the ejecta are divided
and an emission line profile is therefore constructed. 

Given an assumed distance and reddening the original \Nifs\ mass synthesised in
the explosion can be recovered. Using a specific density distribution also allows us
to define the distribution of elements in the ejecta based on reproducing the
line profiles. Strong lines of [\FeII] and [\FeIII] dominate the nebular spectra
of SNe\,Ia and are used for this purpose. Only the inner parts of the ejecta,
which are sufficiently dense at late times and contain enough \Nifs\ can be
studied this way. This complements nicely the early-time spectra, which only
probe the outer layers of the ejecta.

\subsection{Light curve code}

A synthetic bolometric light curve is obtained from the density and abundance
distribution. A Monte Carlo code was described in \citet{Cappellaro97} that first
computes energy deposition, as above, and then follows the diffusion of optical
photons, treating the optical opacity with a simplified scheme based on the
number of effective lines as a function of abundances \citep{MazzaliPodsi06}.
This simplification is justified because it has been shown that line opacity
dominates in a SN\,Ia \citep{Pauldrach96}. A positron opacity of 7\,cm$^2$
g$^{-1}$ was adopted \citep{Cappellaro97}.

\section{Photospheric models}
\label{Best reddening and distance}

Conventionally, the first step in modelling the spectra is to adopt a value for the distance and
reddening of the SN. However, as discussed above, both of these values are
highly uncertain for SN~1991T. Therefore, we include distance and reddening among the fit
parameters.

Another important input is the adopted time of explosion. This is used to rescale the
density profile to the epoch of each spectrum and has a major influence on the
computed synthetic spectra, especially at the earliest times \citep[\eg][]{Mazzali13}. 
Explosion time is referred to the time of the $B$ maximum for which we use the date found by
\cite{1998AJ....115..234L}, JD $2448375.7 \pm 0.5$.

Synthetic spectra were computed varying the distance modulus in steps of 0.1 mag
and, more or less equivalently for small reddening, of 0.03 in reddening (i.e. 10\% steps in
luminosity), while keeping within the range of the measurements given above. We
also varied the rise time by steps of 1 day.
We used values of $\mu$ between 30.30 and 30.76 mags, of $E(B-V)$ between 0.10
and 0.22 mag, and rise times between 18 and 23 days. We found that the best
values for optimal fitting of the spectral series are $\mu=30.57$\,mag,
 $E(B-V)= 0.12$\,mag, rise time $t_r = 20.2$\,days.
The uncertainties on these parameters are about 10\% in the luminosity and one
day in the rise time. From the model it is hard to deduce the intrinsic colour,
hence there is some degeneracy between the best fit values of distance and of extinction.
{The reason for this is the limited ability of our models {to 
 predict accurately} the colour of the spectrum. Especially {post-maximum} spectra 
have too much flux in the red part of the spectrum.}
For example, decreasing the distance modulus by $\sim 0.2$\,mag and increasing
correspondingly the reddening by $\sim 0.06$\,mag we can keep the luminosity of
the model constant and obtain good fits in the photosheric models. However, we
disfavour this solution because the distance would be too low compared to the
values in the literature (Tab. \ref{dist_table}). {On the other
 hand, {a larger distance combined with a smaller reddening} can be
 ruled out {because the} colour of the corresponding modelled spectra 
{is too blue}.}

Subsequently, we experimented with the density profile. We replaced the
``standard'' W7 density profile \citep*{1984ApJ...286..644N} with a delayed
detonation model, which may be more appropriate for a very luminous SN as it can
be more energetic and produce more \Nifs\ than a fast deflagration model like
W7. Keeping the abundances fixed, we only evaluated the effect of the increased
density at high velocity, which is a typical feature of DD models. We selected
model WDD3 from \citet{Iwamoto99} because it is characterised by a large amount
of \Nifs\, ($0.77 \Msun$), which should make it more suitable for a luminous SN
such as SN\,1991T.

\subsection{The early spectra of 1991T}

\begin{figure*}
\begin{center}
     \includegraphics[angle=-90.0, width=1.0\textwidth]{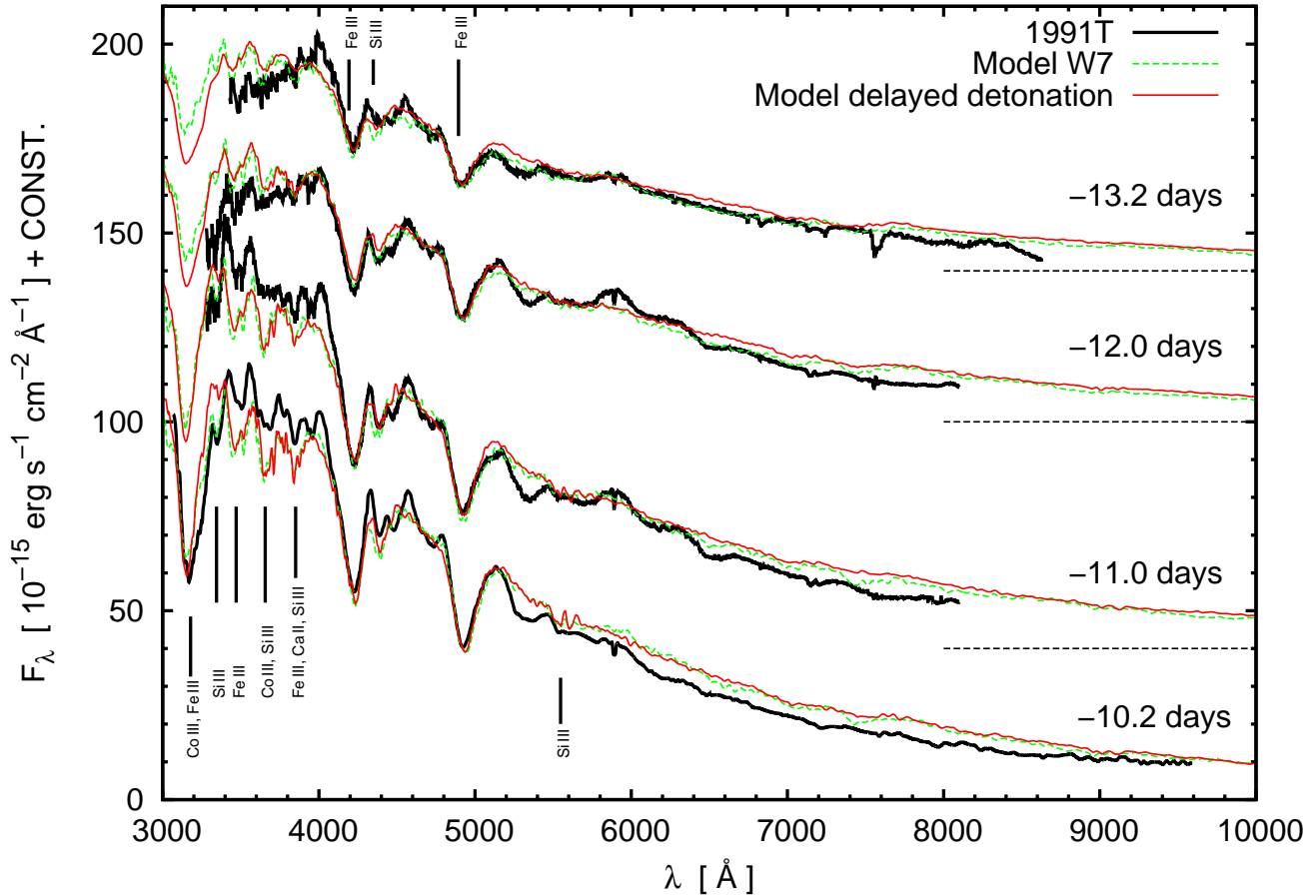}
     \caption{The early spectra of 1991T compared with the best fit models with 
     a W7 and a delayed detonation density profile. The dashed lines on the 
     right represent the zeros values of the spectra.}
     \label{fig: early}
\end{center}
\end{figure*}

In Figure~\ref{fig: early} the spectra from $-13.2$ days to $-10.2$ days since
$B$-maximum are compared with the best fitting models. These early spectra of
SN\,1991T show deep lines of Fe-group elements and only weak lines of IMEs.
Because of the high temperature during this early phase, all the important
features are from doubly ionized species. The species unambiguously identifiable
in the spectra at this epoch are \FeIII, \CoIII, and \SiIII. Both the delayed
detonation model WDD3 and W7 yield good fits of the observed spectra.
The photospheric velocities range between 13000 and 12000\,km\,s$^{-1}$.

\emph{Fe-Group elements.} The most prominent features in these spectra are the
deep \FeIII\ features at 4200\,\AA\ and 4900\,\AA. These lines get deeper with
time as the photosphere recedes inside the ejecta. The models reproduce the
strength and position of these features quite well. Ni, Co, and Fe need to be
present to very high velocity ($\sim 17000$\,\kms) for the model to reproduce
the shape of the spectrum together with the $U$-band magnitudes. Fe-group
elements are rich in near-UV lines, which redistribute the flux to redder
wavelengths \citep{Mazzali2000}. Moreover, \Nifs\ needs to be present in the
outer layers of the ejecta in order to reproduce the deep absorption at
3200\,\AA\ observed in the spectrum at $-10.2$\,days. This sharp feature is
identified as a blend of \CoIII\ and \FeIII\ lines, and Co can only be produced
by \Nifs\ decay. The feature unfortunately is not observed in earlier spectra,
which have a shorter wavelength coverage, and so it is not possible to constrain
more accurately the \Nifs\ abundance at the highest velocities.
{
Modelling a 91T-like SN \cite{2002NewA....7..441H} suggested that at high velocity
there is no need for \Nifs\ because of the absence of \CoIII\ lines. However,
 their models and observations do not study the region within 3100 and 3400\AA\ where 
 we indeed find strong evidence of \CoIII\ lines.
}

Interestingly, some stable Fe is also needed in the outer layers in order to
reproduce the deep \FeIII\ lines in the earliest spectrum, which dates about one
week after the explosion. At such an early epoch \Cofs\ did not have time to
decay to \Fefs. The amount of Fe needed is consistent with solar metallicity.
It most likely originates in the progenitor.

\emph{Calcium.} Most SNe\,Ia show very strong \CaII\ lines (H\&K and IR triplet)
throughout the early phase. SN\,1991T is different. At the earliest epochs, the
IR triplet is practically absent. \CaII\,H\&K lines contribute, together with
\FeIII\ and \SiIII, to a weak feature near 3800\,\AA. \CaII\ lines strengthen only
later. No \CaII\ high-velocity features are observed
 {in the early phases}, unlike in most SNe\,Ia
\citep{Mazzali05}, but this is primarily the effect of temperature. 
{The behaviour of the HVF of Ca {H\&K and IR} at
 later epochs (Section~\ref{subsec:aftermax}), when the feature
appears, can be used to estimate the abundance of Ca between 13000 and 17000~\kms}.
{We set an upper limit on the abundance of Ca at the highest
 velocities that we can investigate where one can expect Ca from the progenitor
 to remain unburned}. {We derive $X$(Ca$)<0.0003$ at $v > 17000$\,\kms. This is
comparable with the solar fraction of this element}. Hence, the Ca in the outer
shells most likely originates from the progenitor. {The mass at
 high velocity ($\sim0.8\times 10^{-4}\Msun$) is a tiny fraction of the total
 mass of Ca, {which is} more abundant at lower velocities.}
{Hence, the uncertainties on the total mass of Ca arise from
 the abundance at lower velocities.}

\emph{Silicon, sulphur, and magnesium.} Among these IME, Si is the only element
that is clearly identifiable in the early spectra. The ionization ratios of Si
are very high compared with normal SNe\,Ia at similar epoch. The \SiIII\ line
near 4400\,\AA\ is easily identifiable in all the spectra. It is due to 
\SiIII\,4553\,\AA, 4568\,\AA, 4575\,\AA. \SiIII\,5740 is weak but still recognizable in this
early epoch near 5600\,\AA. The line disappears from the model using a lower
temperature. On the other hand, in models with an even higher temperature the
\SiIII\,5740 line in the earliest spectrum becomes too strong and does not match
the observations. The \SiII\ line at 6100\,\AA, which is the hallmark of
SNe\,Ia, does not appear at these epochs and this behaviour is correctly
reproduced.

S does not have any clear line at these epochs. Still, the abundance of the
element can be constrained at velocities above 12000\,\kms\ using later spectra.
Sulphur does not show up in the early spectra because it is highly ionized and
there are no strong lines from these species in the observed wavelengths. 

The only strong Mg line in SNe\,Ia is \MgII\,4481\AA\ near 4200\,\AA, but in
SN\,1991T the ionization largely favours the double ionized species, which have
no strong lines, and the feature is dominated by Fe. The amount of Mg is better
constrained by the nebular model.

\emph{Carbon and oxygen} There are no clear C lines in the early spectra of
SN\,1991T. The upper limit for C abundance in the outer layers is
$X$(C)=0.005 and it comes from the appearance of \CII\ lines.
The model can not test the C abundance above 24000\,\kms.
{These correspond to an upper limit of $\sim 0.01\Msun$ of C
 with the WDD3 density.}
 An extensive discussion on the upper limits on the C abundance is
presented in subsection \ref{carbon}. Because of the high ionization, the \OI\
line at 7500\,\AA\ appears only in much later spectra. \OI\ has no strong lines
in the optical. However, by exclusion we conclude that O is the dominant
element in the outer layers. 

\subsection{Spectra before $B$-maximum}

\begin{figure*}
\begin{center}
     \includegraphics[angle=-90.0, width=1.0\textwidth]{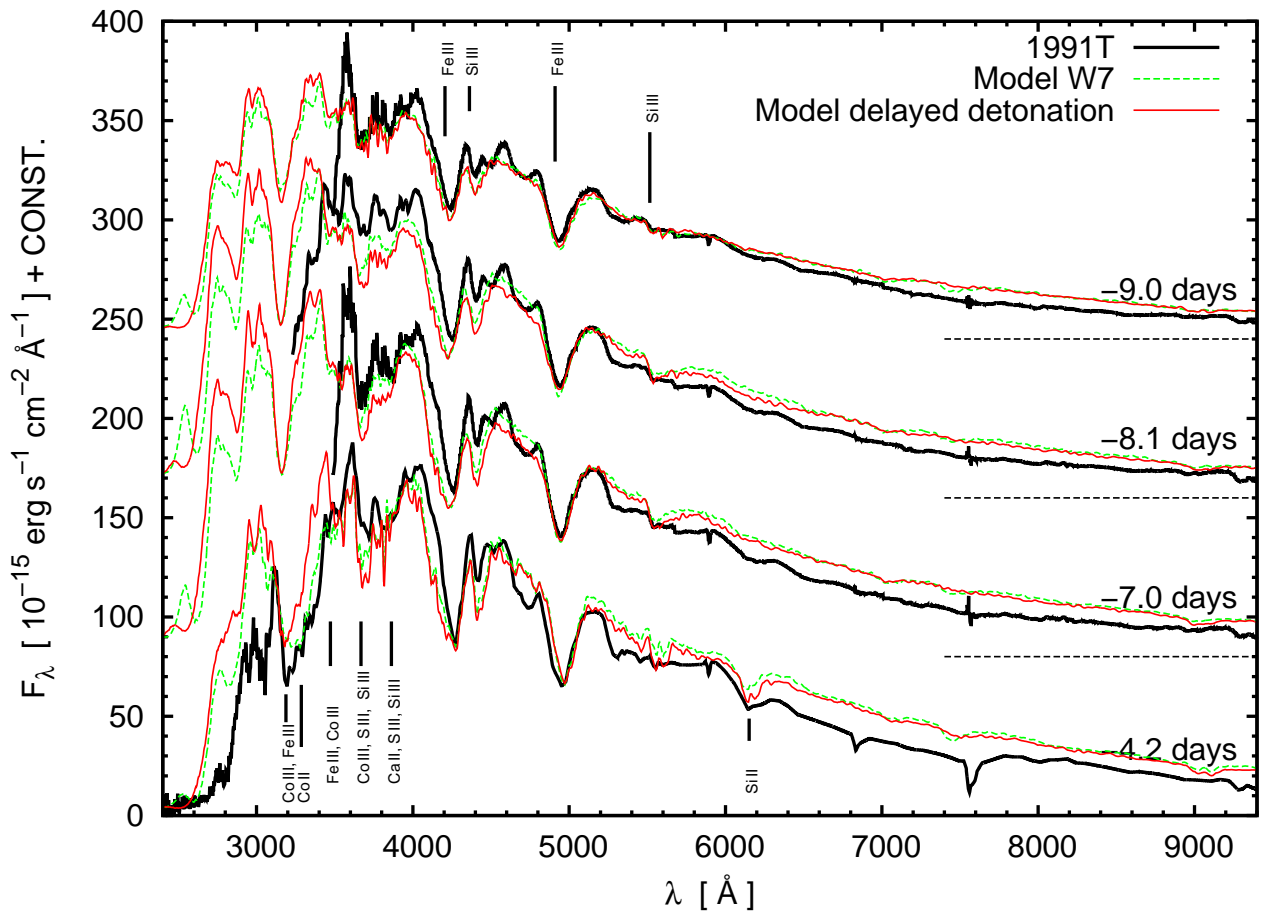}
     \caption{The spectra before maximum of 1991T compared with the best fit 
     models with a W7 and a delayed detonation density profile. The spectrum at 
     $-4.2$~days is connected with an UV spectrum observed from IUE at 
     $-2$~days. The dashed lines on the right represent the zeros values of the 
     spectra.}
     \label{fig: middle}
\end{center}
\end{figure*}

In Figure~\ref{fig: middle} we compare the spectra from $-9.0$ days to $-4.2$
days since $B$ maximum with the best fit model. As in the previous series of
spectra, SN\,1991T shows deep lines from Fe-group elements and only weak lines
from IMEs. Only in the last of these spectra \SiII\,6355 begins to appear.
The photospheric velocities of the model range between 11500 and
10200\,km\,s$^{-1}$.

The observed spectrum at $-4.2$ days is a merging of an optical spectrum and a
UV spectrum obtained by the International Ultraviolet Explorer (IUE) observed 2
days before maximum. IUE obtained three good UV spectra of SN\,1991T between $-2$
days and maximum and they show little variation. We connect the spectrum obtained
at $-2$ days to the nearest available optical spectrum. The UV spectrum has been
rescaled in order to match in the overlapping wavelength range with the optical
spectrum.



\emph{Fe-Group elements.} At the epochs covered in Figure~\ref{fig: middle} the
\FeIII\ features at 4200\,\AA\ and 4900\,\AA\ keep increasing in strength. The
model reproduces this behaviour well. Much more \Nifs\ is required below
12000\,km\,s$^{-1}$ than in the outer layers to block the UV flux. Moreover, 
\CoII\ is required to reproduce the absorption feature at 3200\,\AA. Also, some
stable Fe is required to reproduce the strong \FeIII\ lines.

\emph{Calcium.} The contribution of Ca to the absorption feature at 3800\,\AA\
remains weak. In normal SNe\,Ia the \CaII\ H\&K feature is dominant, but in
SN\,1991T it appears only at later epochs. Also at these lower velocities, there
is no need for a large amount of Ca. 

\emph{Silicon, sulphur, and magnesium.} The ionization degree of these elements
remains very high at these epochs compared with normal SNe\,Ia. \SiII\ 
6355\,\AA\ appears clearly only at $-4.2$ days; instead the \SiIII\ lines at
5500\,\AA\ and especially at 4450\,\AA\ are present at all epochs in this phase.

\SIII\ contributes weakly to the absorption feature at 3800\,\AA, but the
spectra after maximum constrain better the abundances of this element. At $-4.2$
days the \SII\ features at 5300\,\AA\ and 5450\,\AA\ appear for the first time. Although they
are very weak, they are extremely useful in order to follow the shift of the
ionization state of IMEs from doubly to singly ionized species. The model is
able to reproduce the redder of the two lines.

\FeIII\ still dominates the feature at 4200\,\AA, hence it is difficult to
determine the correct amount of Mg from \MgII\ 4481.

\emph{Carbon and oxygen.} At these velocities neither C nor O are necessary to
complete the abundance distribution. At $-4.2$ days a weak \OI\ 7774\,\AA\ line
is produced by the model near 7500\,\AA, but it is not yet present in the
observed spectrum. It appears only after maximum.

Shortly before maximum, the spectrum of SN\,1991T begins to change, and lines of singly
ionized species first appear. This transition gives strong constraints on the model. 
The epoch of the transition is
influenced by the luminosity, i.e., the assumed value of the distance. For a
higher luminosity the temperature increases, making the model worse. This is
not completely reversed by tuning other parameters. Changing the time of 
explosion makes the earliest spectra worse, while a change in the photospheric
velocity ruins the fit of the velocity of many lines.

The model
has too much flux below 3000\,\AA. The mismatch of the model in the UV may be
explained with a possible rapid change in the shape of the UV spectrum during the
time between the two epochs. Many spectral features change rapidly in this
phase. 
This makes a significant change in the UV part of the spectrum
plausible. 
On the other hand, we do not have a way to properly flux calibrate the UV spectrum. 
A calibration problem could also reconcile our model with the observation.
The metallicity of the outer layers can be used to adjust the UV early spectra
 {\citep{2000ApJ...530..966L}} {but reverse fluorescence may dominate this effect \citep{Mazzali2000,2008MNRAS.391.1605S}} leaving largely unaffected the
 visible part of the spectra \citep{Mazzali13}.
We miss early enough UV spectra to accurately constrain the metallicity of the outer layers.

\subsection{The spectra after maximum}

\label{subsec:aftermax}

\begin{figure*}
\begin{center}
     \includegraphics[angle=-90.0, width=1.0\textwidth]{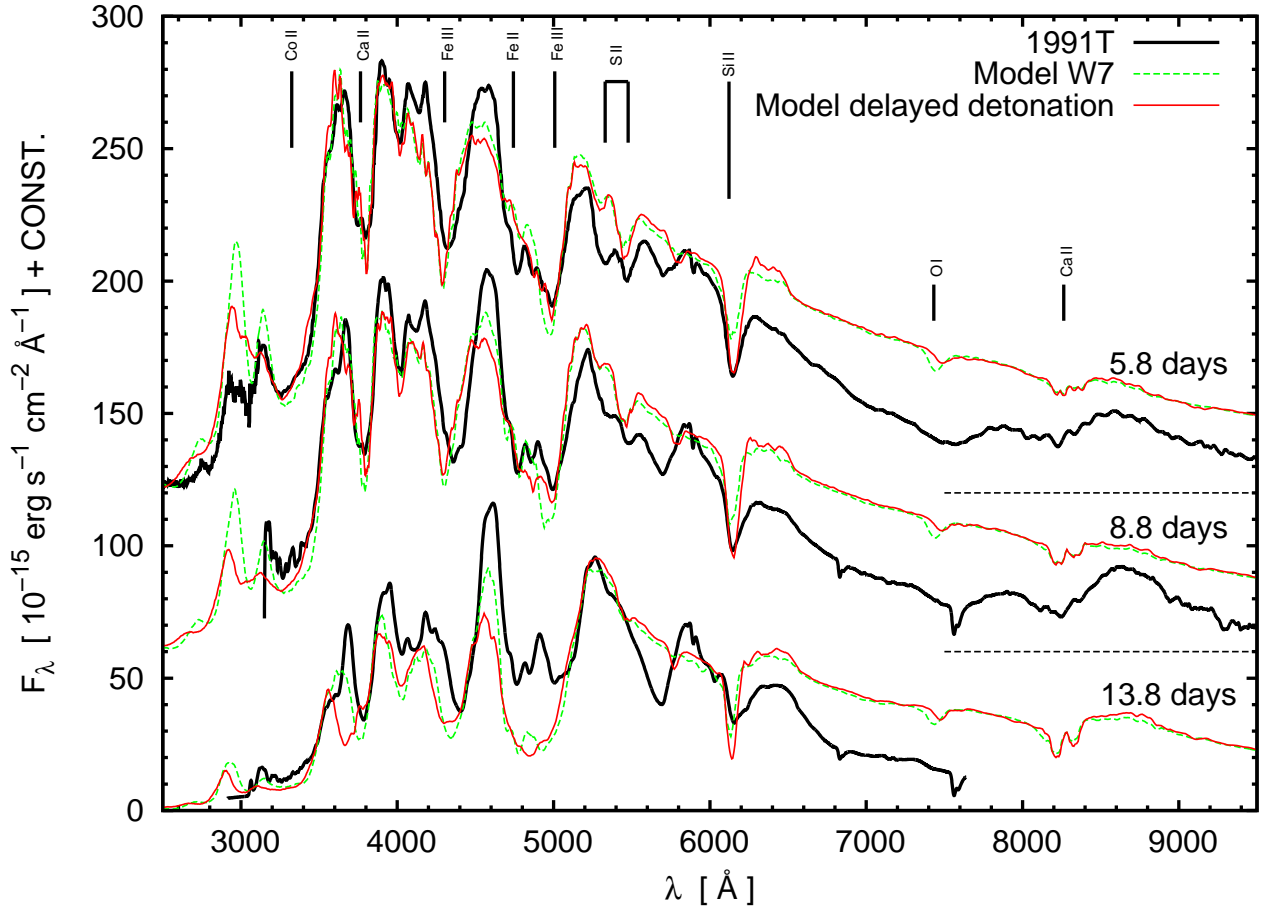}
     \caption{The spectra after maximum of 1991T with the best fit models. 
     The spectrum at $5.8$~days is plotted together with an UV spectrum observed 
     by IUE at $-0.4$~days. The dashed lines on the right represent the zeros 
     values of the spectra.}
     \label{fig: late}
\end{center}
\end{figure*}

Figure~\ref{fig: late} compares spectra from 5.8 to 13.8 days after $B$ maximum
with the best fit models. After maximum the spectrum of SN\,1991T appears
similar to normal SNe\,Ia, with lines of Fe-group elements and of IMEs. The
photospheric velocities of the model range between 9100 and 7200\,\kms. In this
phase, the model overestimates the near-infrared flux because of the failure of
the assumption of black body emission at the photosphere. In a luminous SN\,Ia
like SN\,1991T \Nifs\ is located at high velocities \citep{zorro} and
significant deposition of energy occurs above the model photosphere even a few
days after maximum. Although most line features can be reproduced by our
photospheric-epoch model, the spectra are not nicely reproduced overall, 
especially the later one.
{Although the models frequently overestimate the red tail of the 
spectrum, leading to a larger bolometric flux than observed, we do not expect 
this to cause an overestimate of the ionization state of the matter or to have 
a systematic effect on the occupation numbers of the atomic levels as long as 
the flux in the part of the spectrum that is most strongly coupled with the 
matter (lower than $\sim$5000\,\AA) is correctly reproduced, as is the case in
our models. }
Thus, below $v \sim$9000\,\kms, rather than optimise the abundances of the
photospheric model, we used those derived from the nebular model, which is
sensitive to these low velocities for the same reasons which make our
photospheric model fail.

The spectrum at 5.8 days after $B$ maximum is merged with the nearest available
UV spectrum from the IUE. This was obtained at $B$ maximum, and rescaled in flux
in order to match the part overlapping with the optical spectrum. The shape of
the IUE spectrum does not change significantly over the two days from the
previous spectrum, apart from a decrease in the flux.

\emph{Fe-group elements.} In the nebular model at these velocities \Nifs\ and
some stable Fe largely dominate the abundances. The \FeII, \FeIII\ and \CoII\
lines are well reproduced. If a larger distance is used (meaning a more luminous
supernova) at this
epoch it is impossible to achieve the right ionization ratios. 
The ionization ratio that we mainly use as a proxy for temperature is the
ratio of \FeIII\ and \FeII\ lines, particularly the shape of the broad feature
between 4700 and 5000\,\AA, which is a blend of an \FeII-dominated absorption in
the blue and \FeIII\ lines in the red. Moreover, in models with
{higher} temperatures the strengthening of the \FeII\
features near 4000 and 4200\,\AA\ after maximum occurs later. At these epochs,
 changing the time of explosion does not
influence the spectrum significantly. Alternatively, increasing the velocity of
the photosphere is not sufficient to decrease the temperature of the spectrum,
but it would increase the velocity of the Fe lines, decreasing the quality of
the fit.

\emph{Calcium.} The \CaII\,H\&K lines near 3800\,\AA\ now become important, and
they are reasonably reproduced by the model. Also the \CaII\ IR lines are
present in the spectra and are well reproduced by the model at the correct
velocity, including some of the finer structure, neglecting the offset in flux.
However, it is difficult to constrain strictly the abundance of Ca
 {close to the velocity of the photosphere} because its
ionization favours the double ionized species and it strongly depends on little
changes of other model parameters. 
{The constraints on the Ca at high velocity are stronger because
 a larger fraction is singly ionized. At $5.8$ days the high velocity component
 of \CaII\,H\&K shows as a dip on the blue side of the feature.
}

\emph{Sodium, silicon, sulphur.} The deep feature at 5700\,\AA\ is most likely
to be due to \NaI\,D, but the code largely overestimates the ionization of
sodium and cannot reproduce the line \citep{1997MNRAS.284..151M}.
 At this epoch,
\SII\ features appear clearly for the first time at 5300\,\AA\ and 5500\,\AA.
The \SiII\ line at 6100\,\AA\ is also very important in the spectra. \SiII\ and
\SII\ lines have a higher velocity than the photosphere because doubly ionized
species are dominant at the photosphere. The line strength is difficult to
reproduce using a larger distance because of the higher
ionization ratios, but they are produced at the correct position and with the
correct width (Figure~\ref{fig: late}) with the best fit model.
{The red edge of the \SiII~6355\AA\ line gets shallower with time. This effect is not due to Si but most likely to \FeII\ lines. In particular, in the models at 13.8 days \FeII~6456\AA\ shows an absorption at   
  $\sim$6240\AA. However, in our model the line is not as strong as in the observations.}

\emph{Oxygen} Although O is not present as these velocities, the broad feature
at 7500\,\AA\ may be identified with \OI\,7774\,\AA. The model produces this line
in the outer layers. The line in the models is not broad enough.

\subsection{Summary of the photospheric analysis}

We obtained good fits of a series of spectra covering a range of four weeks
around $B$ maximum using abundance stratification. To fit the spectra we had to
use a lower SN luminosity than most values in literature. Reddening and distance
used in the best fit model are: $E(B-V)=0.12$, $\mu=30.57 $. The best rise time
to $B$ maximum is 20.2 days. These values make SN\,1991T somewhat less luminous
than calculated in, e.g., \citep{Mazzali95}. The freedom in the rise time is
 limited to $\pm1$\,day thanks to the early spectra. 

Compared with models with a higher luminosity there are major improvements in
the ratio between the \SiII\ and \SiIII\ features in the spectra after maximum
(Figure~\ref{fig: late}) and in the last spectrum of Figure~\ref{fig: middle}
($-4.2$ days). A higher luminosity makes the features coming from doubly ionized
species dominant. Also, it is possible to fit the \SiII\ feature in the spectra
after maximum without producing too much \SiIII\ absorption in the early
spectra. Moreover, the \OI\ feature in the spectra after maximum is reasonably
reproduced, albeit with the wrong shape.

The uncertainties in the luminosity of the model and in the estimation of the
 extinction contribute to the uncertainty in the distance.
On the other hand, the intrinsic luminosity of the model has much stronger constraints.
Modelling SN 1991T allows us to have an estimate of the intrinsic luminosity
 of the object independently from the Phillips relation. 
The absolute $B$-magnitude at maximum obtained from the observed light curve
 \citep{2001AandA...368..391R} and from the modelling is M$_B = -19.37$.
An estimate of the uncertainty of this value could be $\pm0.15$ mag.
Due to the nature of the spectra the upper limit on the luminosity of the object is convincing.

An important improvement over \cite*{Mazzali95} is the good agreement with the
UV flux at all epochs, including the earliest U photometry. 

From the spectra in Figures \ref{fig: early}, \ref{fig: middle} and \ref{fig: late}
 it appears that changing the density profile does not lead to significant
changes in the synthetic spectra. Hence, trying different density profiles does
not seem a good way to discriminate among different explosion models on
1991T-like SNe. However, their high luminosity seems to favour energetic
delayed detonations.

\section{Nebular model}
\label{Nebular model}

A late-time spectrum of SN\,1991T obtained on February 5, 1992 with the ESO 3.6m
telescope (+ EFOSC1) \citep{1991ESOC...37..725C} was 
selected for modelling. This date corresponds to 282 days after $B$-maximum,
which is $\sim 302$ days after explosion. We modelled the spectrum using the
NLTE code described in Section~\ref{Nebular_phase}. This spectrum was already
modelled by \citep{zorro}, but the assumptions of distance and reddening were
slightly different in that work, although none of the most important results
change significantly.
We restrict our nebular analysis to this phase to be free from light echo that
 becomes important at phases $>2$yr \citep{1994ApJ...434L..19S}.

We adopt the stratified version of the code and test both explosion models used
for the early-time analysis. Our approach, in the spirit of abundance
tomography, is to use the density distribution as given by the original models
but to modify the abundance of the various elements in order to optimise the
match to the observations. The criteria we use in this process are primarily to
fit the width and flux of the strongest emission lines ([\FeII] and [\FeIII]).

\begin{figure}
\begin{center}
     \includegraphics[angle=0.0, width=0.5\textwidth]{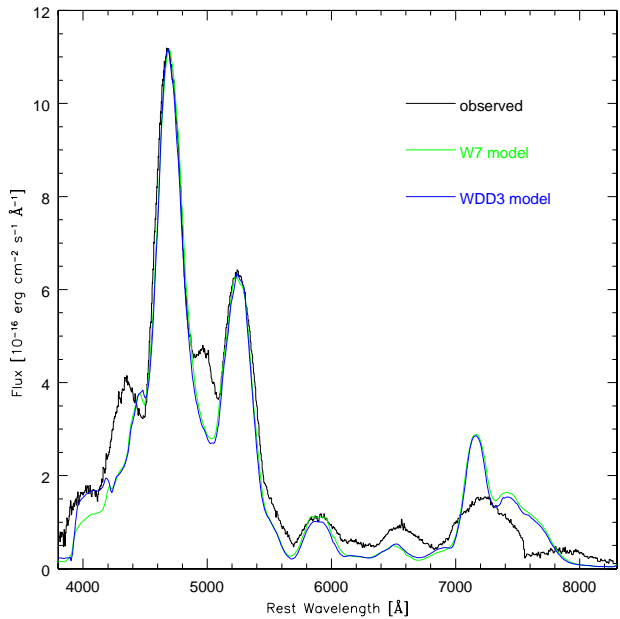}
     \caption{The nebular spectrum of SN\,1991T obtained on Feb 5, 1992 (black)
     compared to two models, obtained using the W7 (green) and WDD3 (blue)
     densities.
}
     \label{nebular}
\end{center}
\end{figure}

We are able to obtain reasonably good fits for both density stratification
models, with only small difference between them (Figure~\ref{nebular}).
In the case of W7 we find a \Nifs\ mass of $0.78 \Msun$. \Nifs\ is located
between 2000 and 13000\,\kms, and is dominant between 4000 and 12000\,\kms. The
large extension of the \Nifs\ zone explains the width of the nebular emission
lines, which are among the broadest for any SNe\,Ia \citep{Mazzali98}. The inner
layers are dominated by stable \Feff, whose mass we estimate to be $0.27 \Msun$.
This is necessary in order to match the flux and shape of the lines, as well as
to reach a reasonable ionization degree. The total production of NSE material is
$\sim 1.16 \Msun$. 

The delayed detonation model yields the same \Nifs\ mass, $0.78 \Msun$. The
distribution with velocity is also similar. The DD model contains less mass in
the lowest velocity shells, and this can be compensated by using less stable Fe
in these zones, so the mass of \Feff\ is now only $0.15 \Msun$ and the total NSE
mass is $0.94 \Msun$. In both cases the mass of \Nife\ is small ( $\sim 0.01 \Msun$), as set by the
non-detection of the line at 7378\,\AA.

The two fits are similar in quality, which makes it impossible to differentiate
between these two models based on nebular spectroscopy. However, if we adopt the
W7 density distribution, the abundances are not compatible with those of W7,
while in the case of WDD3 they are closer to those of the original model
\citep{Iwamoto99}. This makes us favour the DD model.

\section{Abundances of elements}
\label{Abundances of elements}

\begin{figure*}
\begin{center}
     \includegraphics[angle=-90.0, width=1.0\textwidth]{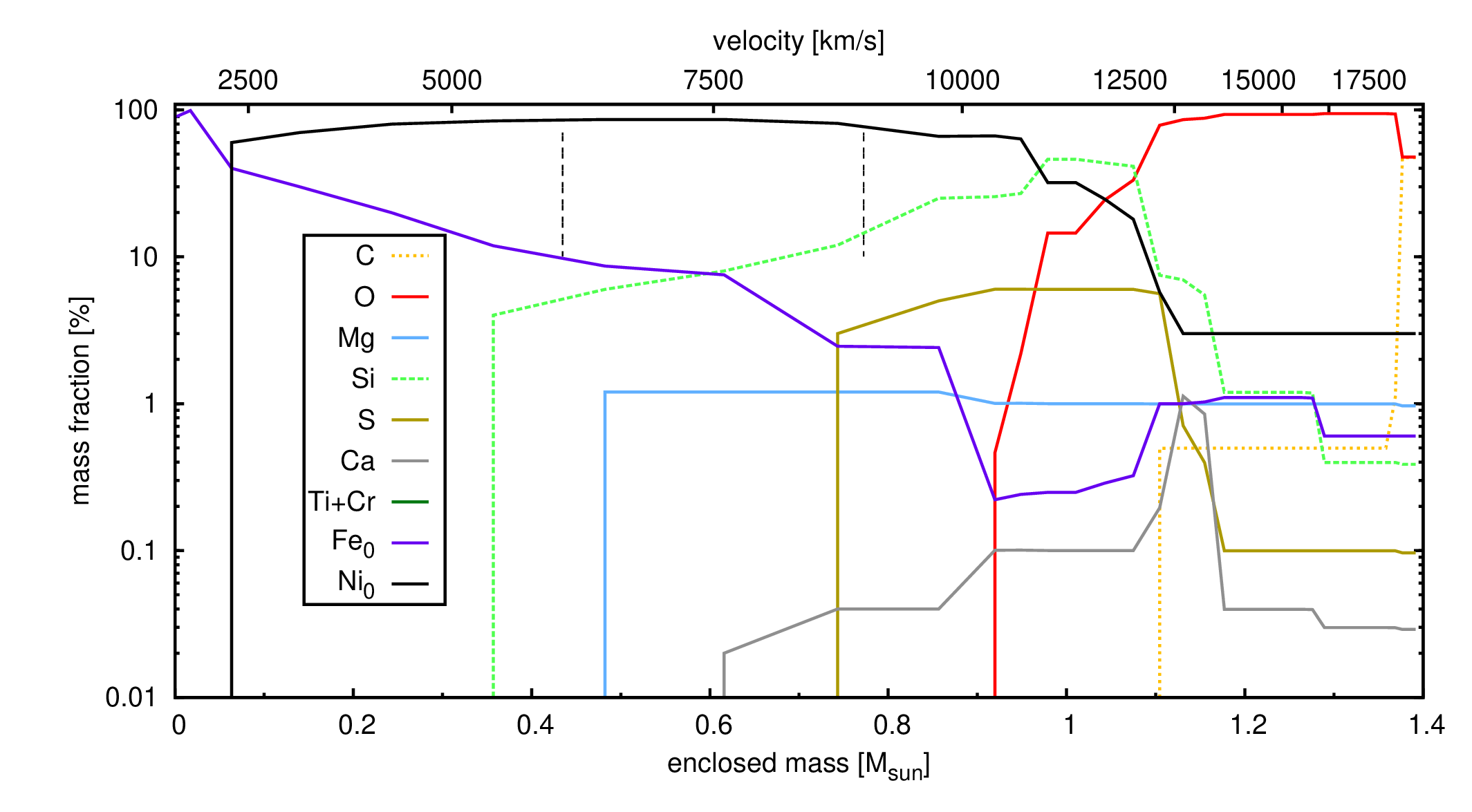}
     \includegraphics[angle=-90.0, width=1.0\textwidth]{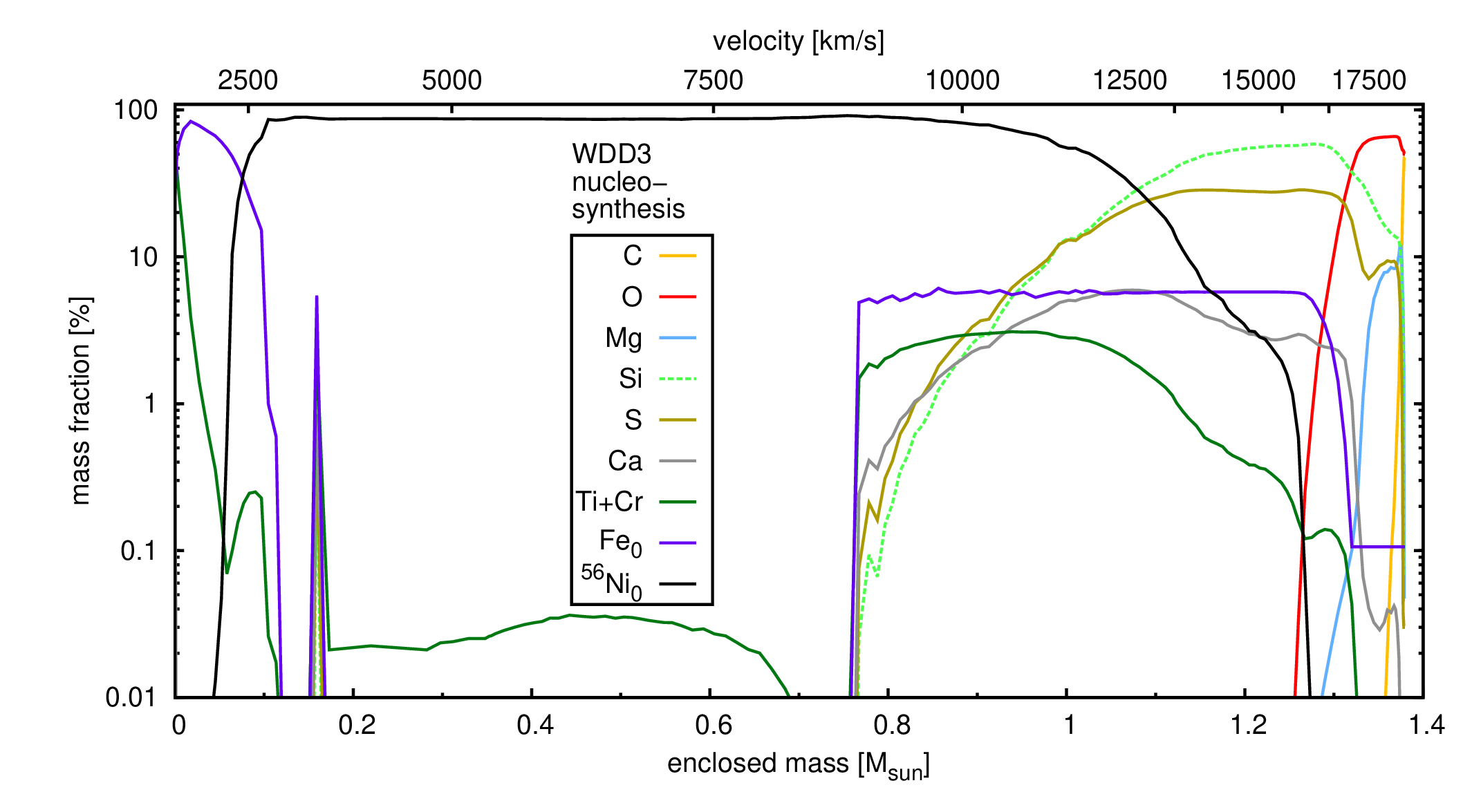}
     \caption{Abundances of the elements of the best fit model with the delayed 
     detonation density profile of the SN\~1991T (top panel, plotted in 
     mass space).
 The abundances as a function of velocity are not significantly different between this model 
 (obtained with the WDD3 density profile) and the one with the density profile 
 of the W7 model (see text, and Figure~\ref{abundances_w7_1991T}). 
{The vertical dashed lines correspond to 6000 and 9000 \kms.
 The abundances in this region are computed by the nebular model and used by
 the photospheric models without significant changes.}
The C abundance represented by the dotted line is the upper limit allowed in 
 the model.
The bottom panel represents the abundances of the WDD3 nucleosynthesis 
calculation.
 The major difference is the absence in our model of a Si dominated shell.}
          \label{abundances1991T}
\end{center}
\end{figure*}

Figure~\ref{abundances1991T} shows the abundances over radius in the best-fitting
model with the delayed detonation density profile. Our model is compared with
 the original abundances. At velocities below
6000\,\kms\ the abundances are from the nebular model. Between 6000\,\kms\ and
9000\,\kms\ the abundances from the nebular model are used in the photospheric
models without changes, above 9000\,\kms\ the abundances are from spectral
models with the photospheric code.

\begin{figure*}
\begin{center}
     \includegraphics[angle=-90.0, width=1.0\textwidth]{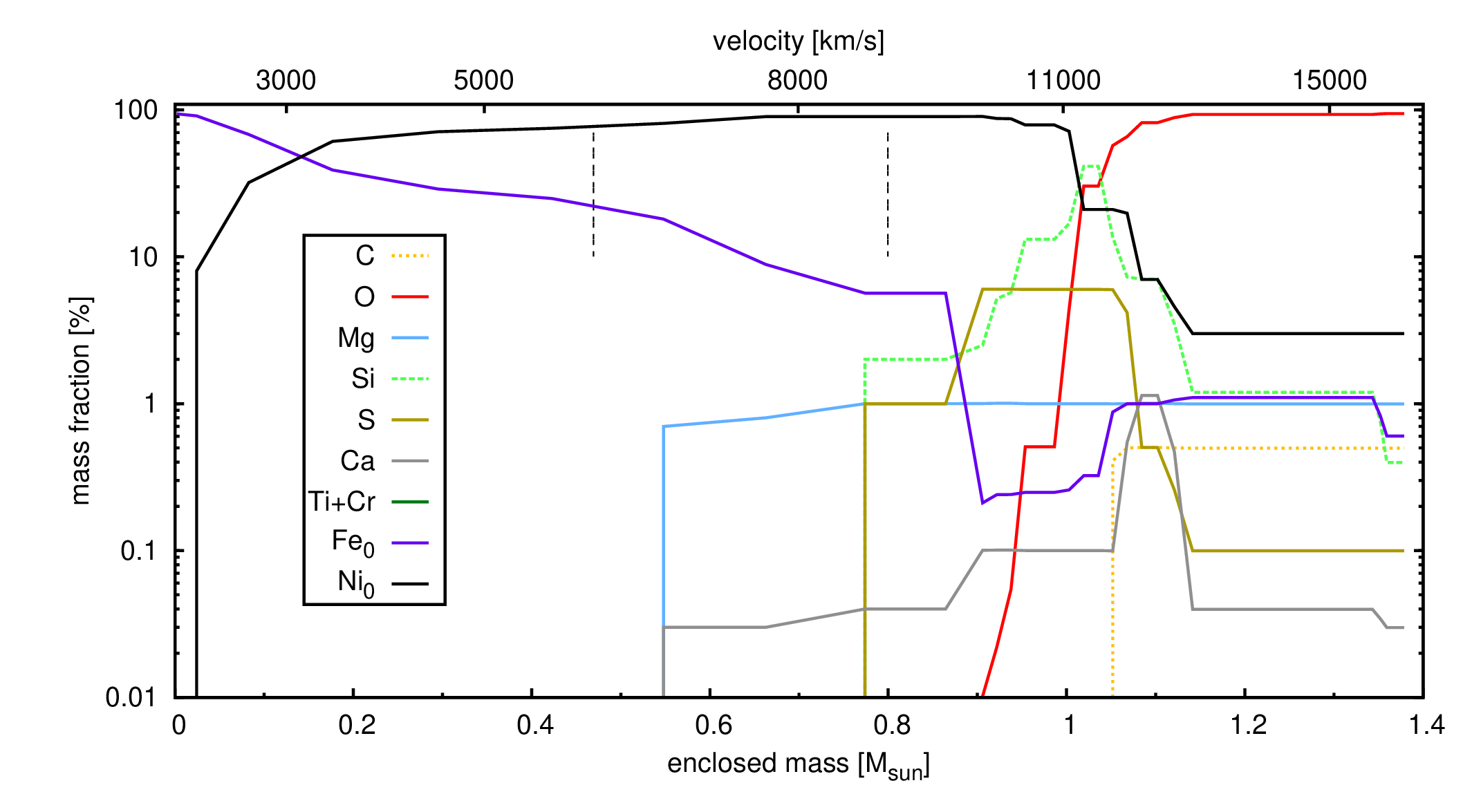}
     \includegraphics[angle=-90.0, width=1.0\textwidth]{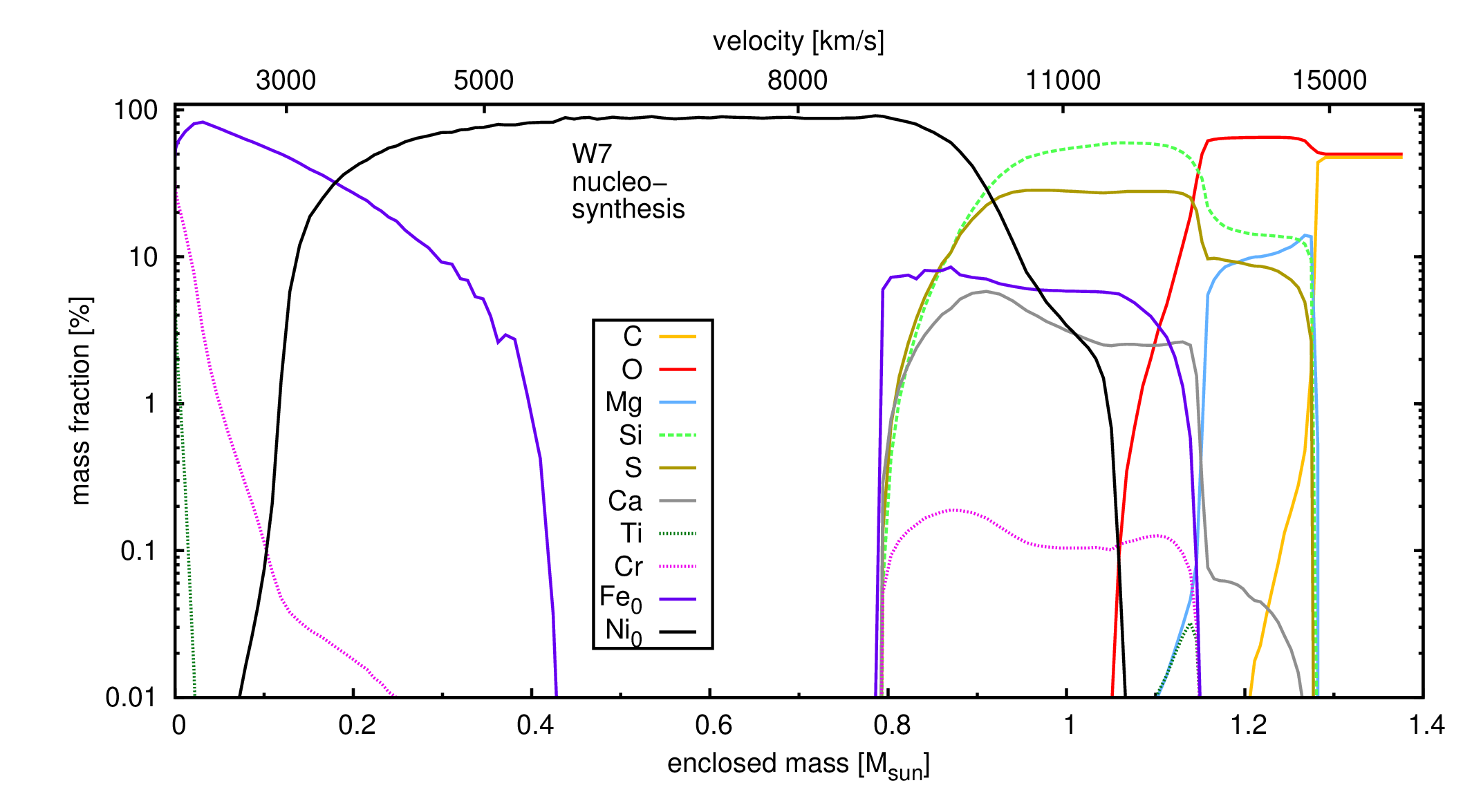}
     \caption{Abundances of the elements of the best fit model with the W7
     density profile of the SN\~1991T (top panel, plotted in 
     mass space).
  The C abundance represented by the dotted line is the upper limit allowed in 
  the model.
The bottom panel represents the abundances of the W7 nucleosynthesis 
calculation.
  The major differences are the larger amount of \Nifs {and stable Fe},
 the scarcity of IMEs and a large shell with O but no C.
  {The major differences with the model based on the WDD3 density 
 (Figure~\ref{abundances1991T}) are more stable Fe, less IMEs and a stricter upper limit on C.} 
}
          \label{abundances_w7_1991T}
\end{center}
\end{figure*}

\begin{figure*}
\begin{center}
     \includegraphics[angle=-90.0, width=1.0\textwidth]{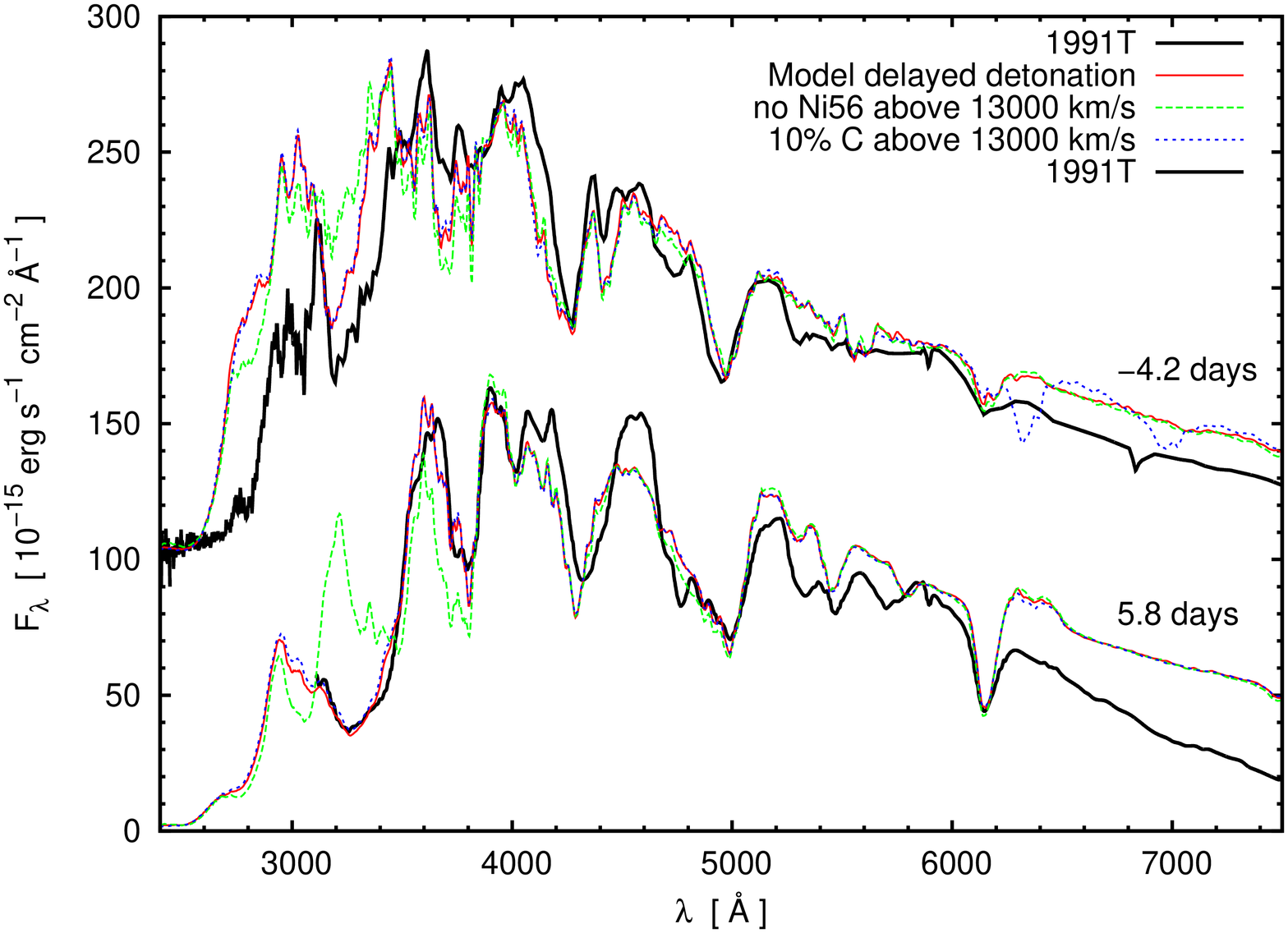}
     \caption{ The spectra of 1991T at two significant epochs are compared with the 
               delayed detonation density model.
              The effect of removing the high velocity \Nifs\ or of adding a large 
              quantity of C on the delayed detonation density model
              are shown to worsen the quality of the fit.
                }
          \label{out_Ni_C}
\end{center}
\end{figure*}

Our experience from computing many different models for the photospheric phase
 is that the stratification
of abundances is quite fixed, but the whole picture can shift in velocity by
about 1000\,\kms\ depending on the rise time. This makes the measured masses of
\Nifs\ (0.78M$_{\odot}$) and of O (0.29M$_{\odot}$) not very precise using the
 photospheric model alone. For
example, reducing the rise time by half a day reduces
the mass of O by $0.1\Msun$, increasing the mass of the Fe and \Nifs\ rich core
while maintaining the photospheric spectra almost unchanged. On the other hand,
the nebular model constrains precisely the mass of \Nifs\ 
(section \ref{Nebular model}). The results of the nebular model are affected by
the distance to the supernova. The uncertainties in this parameter will
propagate to the estimate of the \Nifs\ mass. Hence, a safe estimate of the
error over the Ni mass could be 0.05M$_{\odot}$.

The photospheric models need a significant amount of iron-group elements ($\sim 4$\%) at high
velocity, above 12000\,\kms. In particular about $3\%$ of \Nifs\ needs to be at
 high velocity to
reproduce the broad and deep absorption feature comprised between 3200\,\AA\  
and 3400\,\AA. The time evolution of the feature is well reproduced by the
model. The main contributors to the feature are a large number of \CoIII\ lines 
in the spectrum at $-10.2$ days, both \CoIII\ and \CoII\ lines at $-4.2$ days
and by \CoII\ lines afterwards. The \CoII\ lines contributing to the absorption
feature have lower-lying levels, with energies mostly comprised in the 2-3eV
range. Removing \Nifs\ from the outer shells makes it impossible to reproduce
the absorption feature, as is shown in Figure~\ref{out_Ni_C} where the model
with the delayed detonation density profile is shown with and without the high
velocity \Nifs. This is a refinement over the modelling from \cite{Mazzali95}, who
claimed that SN\,1991T was dominated by iron-group element also in the outer
layers. That result came from requiring an excessively high luminosity because of
the assumption of a larger distance than adopted here. We should stress again
that the lower distance is preferred because it yields generally improved fits
to the spectra at all photospheric epochs. Lowering the ionization actually
increases the intensity of \FeIII\ lines in the spectra given the same
abundances. This lower luminosity proved crucial to model many other
characteristics of the spectra. 

The presence of \Nifs\ in the outer layers has some of the characteristics of a high 
velocity feature \citep[HVF][]{2005ApJ...623L..37M}. This \Nifs\ is placed at
 a velocity much higher than the majority of the elements. The high velocity
 shell is devoid of C and made mostly by O. A similar behaviour is typically
 shown also by the Ca HVFs.

The abundances of other elements are also measured by the model with some
accuracy. In Figure~\ref{out_Ni_C} a model with 10\% of C in the outer shell is
also shown. This is much more than what can be allowed to reproduce the spectra.
At both epochs the ionization ratio favours the \CII\ species at high velocity.
In the C-enriched model \CII\ features at 6370\AA\ and at 7000\AA\ are produced
by lines with lower levels at 14 and 16eV, respectively. The occupation numbers
of these levels drop in the spectra after maximum because of the lower
temperature. Instead the occupation levels are optimal to detect the line in
pre-maximum spectra. On the other hand, the increasing temperature of the
radiation field at early epochs (earlier than a week before maximum) favours
species of C with a higher ionization also in the outer layers. It is possible
to obtain a strict upper limit on the amount of C in the supernova using the
spectra before maximum. C must be less than 0.5\% in the O-rich shell, because
the \CII\ lines that appear already with this quantity of C are not present in
any of the observed spectra.

The amount of Mg is quite uncertain, but it must be less than 20\% in the O-rich
shell, because with this amount the line at 4300\,\AA\ becomes too deep owing to
the contribution of \MgII\ 4481\AA. 

The amount of Si in the best-fitting model is about $0.14 \Msun$. The uncertainty in
this result can be estimated as $0.05 \Msun$. This is large because the only
really useful line is \SiIII~4450\,\AA. The \SiII~6355\AA\ line in the
post-maximum spectra is sensitive to the \SiII/\SiIII\ ionization ratio, hence
the actual luminosity of the model affects this line more than the abundance of
Si itself. This is the same behaviour as the \SiII~5972\AA\ line in normal
SNe\,Ia \citep{Hachinger08}. However, silicon, the
most abundant IME, has a similar distribution in velocity also in models with
larger luminosity. The shell where IMEs are most abundant is very narrow,
between 10000\,\kms\ and 11500\,\kms, and both \Nifs\ and O are present with
significant abundances even in this shell. The narrow IME zone is the main
difference between SN\,1991T and other SNe\,Ia. The total amount of IMEs in the
model is about $0.18 \Msun$.

The abundance of Ca is determined from \CaII\ H\&K. {At velocities
 lower than $~13000$\kms, where the bulk of Ca is present,} also this line is more
sensitive to the ionization state than to the actual abundance of the element. The
position of the line shows that this element must peak at a higher velocity than
Si and S. The velocity range enriched in Ca is between 12300\,\kms\ and
12500\,\kms. This is unexpected, because Ca is produced in explosive
nucleosynthesis after Si and S, and the combustion progresses further to NSE
going deeper into the star.
{
Moreover, the ejecta show clear signs of \Nifs\ at even higher velocities.
It looks like part of the ejecta underwent a more complete combustion in the outside
 and a progressively limited combustion in the inside. This is in contrast with the bulk
 of the ejecta, showing the usual stratification.
}.
 This velocity ``inversion'' between {\Nifs,} the Ca-rich
layer, and the layer where the other IMEs are dominant can be an important hint
to understand the underlying explosion mechanism.
{
This could be explained by ``plumes'' of material with more complete burning
 expelled at velocities higher than the bulk of the ejecta.
Another possible explanation could be the detonation of a thin He layer present
 in the outer part of the white dwarf. In the framework of the delayed detonation
 models, this layer may get triggered by the detonation front.
}
 The total amount of Ca found
by the model is about $0.9\times10^{-3} \Msun$. On the other hand, it is known that
high-velocity \CaII\ lines exist in most SNe\,Ia, and these may be the result of
the interaction of SN ejecta with H-rich circumstellar material
\citep{Gerardy04,Mazzali05,Tanaka08}.

{Figure~\ref{abundances_w7_1991T} shows the abundances for the
 best-fitting W7 density profile. 
The overall picture is similar to the results of the WDD3 density model.
However, there are significant differences in the centre. With the W7 model we
 obtain a larger amount of stable Fe in the centre (Section~\ref{Nebular model}).
 The luminosity constrains the mass of \Nifs\ to be similar in the two nebular models. 
The larger mass of Fe-group elements in W7 is compensated by a lower mass of Si and
 other IMEs at intermediate velocities.
Unfortunately, the amount of IMEs is {difficult to test} using nebular analysis.
Velocities lower than $\sim$10000\kms\ are also difficult to assess via 
photospheric analysis because, although the velocity of the photosphere drops
 below this value at maximum, the ionization state of Si and S close to the
 photosphere is too high to affect the spectra. 
For these reason with the W7 density profile we obtain a similar {composition} as with the WDD3 density, apart from a larger amount of stable
 Fe ($\sim$$0.27\Msun$) and correspondingly, a lower amount of IMEs ($\sim$$0.04\Msun$).
 }

\subsection{Carbon}
\label{carbon}

Evaluating the amount of unburned material is very important to constrain
different explosion models. We can estimate the amount of unburned material
by measuring the abundance of C allowed in the model. This element must belong
completely to the original composition of the CO white dwarf because, unlike O,
it can only be destroyed by nuclear burning in an environment devoid of lighter
elements.

There are no clear C lines in the spectra. In order to place an upper limit to
the amount of this element in our model, we assume two different types of
distribution for the element: a uniform abundance in the O-dominated shell, or
all C being confined in an outer, unburned shell, which we assume to be 
composed of C and O in equal amounts.

\begin{figure}
\begin{center}
     \includegraphics[angle=-90.0, width=0.5\textwidth]{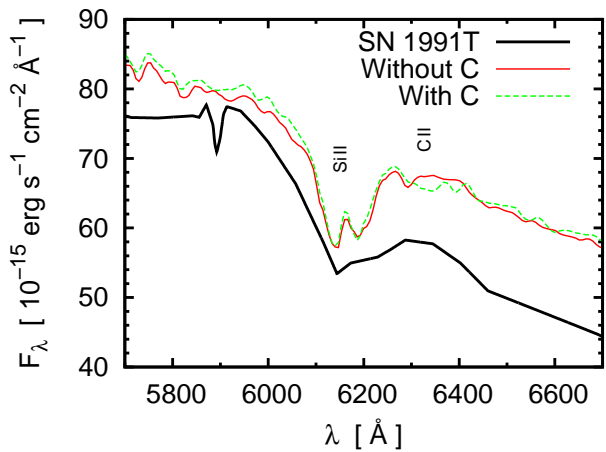}
     \caption{In the Figure we compare the model with the amount of C in its 
     upper limit with the model without this element.
  Only the section of the spectrum at $-4.2$ days affected by a 
  small amount of C is shown. }
     \label{carbon 1991T}
\end{center}
\end{figure}

The synthetic spectrum that is most affected by the uniform distribution of C in
the O-rich shell is that at $-4.2$ days, because of the lower ionization which
allows a \CII\ line at 6300\AA\ to appear. In Figure~\ref{carbon 1991T} we
compare two models, with and without C. Both models are based on the delayed
detonation density profile. We set an upper limit of 0.5\% to the C abundance,
because for lower values it is no longer possible to distinguish the C line;
moreover the spectrum resolution is not accurate enough to sustain a tighter lower limit.

The spectrum that is most influenced by introducing C at high velocity is the
earliest one. Since SN\,1991T was observed from quite early on and the first
spectrum shows no clear evidence of C, the lower limit for the velocity of a
shell of unburned material is quite high, that is 24000\,\kms.
The recently discovered, spectroscopically normal SN\,2013dy shows strong
 \CII\ lines detached from the photosphere in the early epochs \citep{2013ApJ...778L..15Z}.
The velocity of these HVFs of C evolves to a minimum of $\sim23000$\,\kms.
Our model shows that if C is present in the peculiar SN\,1991T it has to be at
 an even higher velocity.

Doubly ionized C is not useful in determining the upper limit of the C abundance.
\CIII\ lines appear in the spectra only with abundances high enough to produce
 strong \CII\ lines.

Combining the two upper limits, the model allows a maximum amount of C of
0.01M$_{\odot}$, at the expense of O, using the delayed detonation density
profile.

With the W7 density, the upper limit to the amount of C is much lower 
($0.001\Msun$). The main reason is the much lower amount of material at
high velocity, above $\sim$\,$22000$\,\kms\ in the W7 model. The C in this
high velocity material does not show up in the spectra with either the W7 or
the delayed detonation density profile also with a 100\% abundance. With the
 W7 density the amount of C that can remain
``hidden'' at such high velocities is negligible. 

However, it is difficult to believe that there is no mixing of material at all
at these very high velocities and that the interface between O and unburned
material is so sharp. If there was some mixing or a smooth transition between
these two zones, the maximum C mass allowed would be even lower. This is because a
significant amount of C between 15000 and 24000\,\kms\ would be easily
detectable. At the same time, to have a smooth transition the amount of C above
24000\,\kms\ has to decrease significantly.

\section{Bolometric light curve}
\label{Bolometric_light_curve}

\begin{figure}
\begin{center}
     \includegraphics[angle=0.0, width=0.5\textwidth]{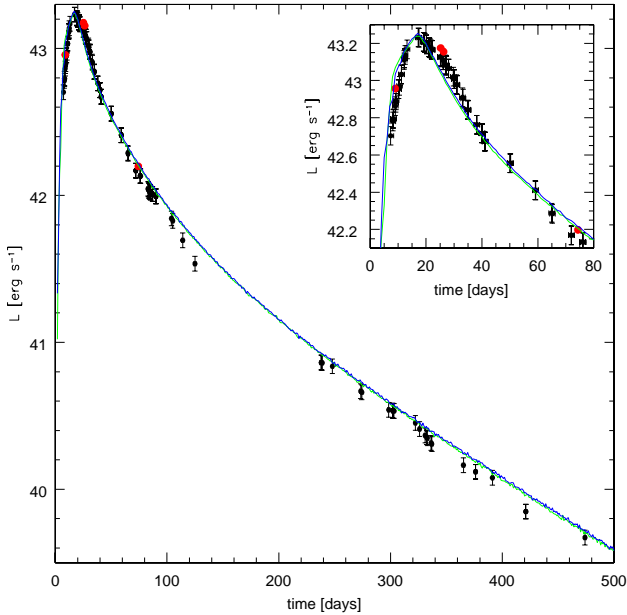}
     \caption{Synthetic bolometric light curves based on the
     results obtained using the density structures and the abundance
     distributions derived for W7 (green) and WDD3 (blue) compared to the
     $UBVRI$ pseudo-bolometric light curve of SN\,1991T (black points). The
     correction for the infrared flux is shown as red points for the few epochs
     in which it was possible to derive it based on the published data.  
}
     \label{LC_figure}
\end{center}
\end{figure}

To check the consistency of the model we compare the bolometric light curve with
the observed magnitudes. We have calculated a bolometric light curve from the
$UBVRI$ photometry. For each epoch of observation (see Table in Appendix
\ref{mag_table}, data from \citealt{2004MNRAS.349.1344A, 1998AJ....115..234L}),
we constructed a spectral flux distribution using the flux zero-points of
\cite{1995PASP..107..945F}. For epochs when observations are missing in one or
more filters, we interpolated the magnitudes using splines of the light curves. 
Since the light curves are well sampled and regular, this introduces a
negligible uncertainty. The spectral flux distributions were dereddened with
the extinction curve of \cite{1989ApJ...345..245C}, using $E_{B-V}$ = 0.12,
splined and integrated in the wavelength range 3000-10000 \AA\ (a linear
extrapolation of the flux was used bluewards of the limit of the $U$-band filter
and redwards of the limit of the $I$-band filter). Using the distance modulus
$\mu = 30.57$ we derived bolometric luminosities, which are shown in Figure~\ref{LC_figure}.
The near-infrared contribution to the {observed} bolometric
 light curve is difficult to
estimate, because observations are only available at four epochs
\citep{2004AJ....128.3034K}. {Using the observations, the near-infrared
luminosities at these four epochs are estimated to be 14\%, 3.3\%, 3.5\%, and 6.4\%
 of the total luminosities
in the optical and near-infrared.} These contributions are small, and
comparable to the uncertainty in the derived bolometric luminosity.

We produced synthetic bolometric light curves using the code described in
Section~\ref{Method} and the abundances derived from spectral modelling.
 In the spirit of
abundance tomography, this is a test that the derived abundance (and density)
distribution is appropriate for the SN at hand. We used both the W7 and WDD3
densities and the abundance stratification derived from the respective
tomography experiments.  The results are shown in Figure~\ref{LC_figure} 
 and compared to the light curve of SN\,1991T.  
{From the synthetic spectra we can set an upper limit of
10\% for the NIR contribution to the total spectral flux at all epochs where
 spectra were computed. Based on these considerations, we can safely neglect the NIR contribution to
the bolometric luminosity.} The agreement
is satisfactory in both cases.
 Bolometric maximum is reached about 18 days after
explosion with both models. The differences between the two model light curves
 are smaller than the uncertainties in the construction of
the bolometric light curve and those deriving from the use of a constant opacity in the model. At
later times the two light curves are basically indistinguishable, because the
two models have the same \Nifs\ mass.

\section{consistency of the model}
\label{Check_the_consistency_of_the_model}

The amount of \Nifs\ in our best fit model with the delayed detonation density
 profile for SN\,1991T is $\sim 0.78 \Msun$. Stable Fe
is about 0.15 M$_{\odot}$, there are some IMEs (only $\sim 0.18 \Msun$), the
rest is supposed to be O (0.29 M$_{\odot}$) while essentially no C is present. 

In the model with the W7 density profile the amount of stable Fe is larger,
 $\sim 0.27$. The quantity of \Nifs\ is similar, $\sim 0.78 \Msun$.
 The \Nifs\ mass correlates strongly with the luminosity in the nebular phase.
The mass of the IMEs is smaller (only $\sim 0.04 \Msun$) and the amount
 of O is similar ($\sim 0.30 \Msun$).

The maximum bolometric luminosity computed from the photometry
with the distance modulus and reddening used here is $1.7\times 10^{43}$\ergs.
 Using this luminosity together with a version of the Arnett's rule
 \citep{1982ApJ...253..785A} from \cite{2006A&A...460..793S} 
 we derive for SN\,1991T $\sim 0.85 \Msun$ of \Nifs,
with an error of about $\pm 15$\%. This is consistent with the
value we derived (0.78M$_{\odot}$).

\begin{table}
\caption{Nucleosynthesis }
\begin{tabular}{lcc}
\label{abund_table}

 & W7 density & WDD3 density \\
\hline
O                  & $0.30\Msun$ & $0.29\Msun$  \\
IME                & $0.04\Msun$ & $0.18\Msun$ \\
\Nifs\             & $0.78\Msun$ & $0.78\Msun$ \\
stable Fe          & $0.27\Msun$ & $0.15\Msun$  \\
$E_K^{\text{ from nucl.}}$ & $1.27\times 10^{51}$erg& $1.24\times 10^{51}$erg \\
$E_K$              & $1.30 \times 10^{51}$\,erg& $1.43 \times 10^{51}$\,erg\\

\hline
\end{tabular}
\end{table}

Calculating the kinetic energy from the nucleosynthesis using the formula:
\[ E_K=[1.56M_{\text{\Nifs}}+1.74M_{\text{stable Fe}}+1.24M_{\text{IME}}-0.46]\times 10^{51} \text{erg} \]  
from \cite{2007ApJ...662..487W} {gives for the model with the WDD3
density $1.24\times 10^{51}$erg and for the model with the W7 density a similar
value of $1.27\times 10^{51}$erg. This only is about 3\% less than the kinetic
energy of the W7 model \citep[($\sim 1.30 \times 10^{51}$\,erg,  ][]{Iwamoto99}. 
It is also smaller than the kinetic energy of the WDD3 model by about
13\%  \citep[$1.43 \times 10^{51}$\,erg, ][]{Iwamoto99}.} The low energy is 
caused by the lack of IMEs. 
{These results are reviewed in Table \ref{abund_table}}.
{The energy released by burning a large amount of material
 to Fe-group elements in SN\,1991T is not enough to compensate for the large
 amount of energy which remains untapped in the unburned oxygen layer.}
 However, as 
explained above, a model with a rise time shorter by half a day can reproduce 
the spectra almost as well and requires a smaller amount of O.  In that case the 
energy computed from the nucleosynthesis is about $1.3 \times 10^{51}$\,erg.

We find that it is possible to achieve good fits only with a lower luminosity
than what some of the distance estimates imply. Measures of distance to the host
galaxy of SN\,1991T are quite
different (Tab. \ref{dist_table}),
and the distance modulus used here is compatible with most of these
measurements. Moreover, it is difficult to distinguish our model from one with a
distance modulus increased by 0.1\,mag and reddening decreased by 0.03 mag,
because these two parameter have a similar effect on the spectra. 
The uncertainty on the extinction is larger than that on the distance.

{  In this work we used the density profiles as given. Among all 
elements used to fit the spectra, O has the smallest direct effects on the
spectra. Relaxing the constraint on the density profile can improve the
energetic consistence of the fit. It is possible to remove some of the oxygen
while still preserving good spectral fits in the early epochs. This means
removing up to $\sim 0.1 \Msun$ of O from the external layer. However, although
O in the outer layers does not produce strong lines (with the exception of the
\OI\ line at 7500\,\AA), its presence increases the electron density. This is
important for the ionization balance of the other elements. Removing O increases
the ionization,  {and a smaller amount of Fe is needed above
 $\sim 16000$\,\kms than with the W7 or the WDD3 densities. Differently from
 what happens with these density profiles,} the Fe from \Cofs\ 
decay is enough to reproduce the \FeIII\ lines in the pre-maximum epochs.


The presence of less material at high velocity is also suggested by the kinetic 
energy balance. The density profiles used here have too much kinetic energy 
compared to the nucleosynthesis. The kinetic energy can be significantly  
reduced by modifying the outer layers. It is important to keep in mind that the
balance of the kinetic energy assumes a Chandrasekhar mass white dwarf. This 
discussion does not necessarily mean that we suggest a mass lower than 
$1.4\Msun$. Most likely, the mass subtracted from the outer layers can be 
redistributed in the much more massive inner parts without too much changes in 
the spectra. A low density layer at high velocities, composed of \Nifs\ and Ca 
could be an important clue of the explosion mechanism.

On the other hand, a low density at high velocity with a significantly lower 
kinetic energy could be an indication of a Super-Chandrasekhar mass ejecta. This 
would also naturally explain the abundance of stable iron in the centre inferred 
from nebular spectroscopy. The slow evolution of the light curve is also 
favoured by a larger total mass.  In any case the deviation from the 
Chandrasekhar mass should not exceed 10\%.

Changing the density profile of the ejecta to explore these possibilities should 
be done using some supporting models that assure hydrodynamic consistency. 
Otherwise there may be too many free parameters to be investigated and there is 
the risk to end up with an unphysical density profile. This scenario deserves 
careful investigation, but this exceeds the aim of this work. }

\section{Comparison with explosion models}
\label{Comparison with explosion models}

The distribution of elements rules out strong mixing as predicted by 
three-dimensional deflagration models \citep[e.g. ][]{2005ApJ...623..337G,
2005AA...431..635R, 2007ApJ...668.1132R}.
Delayed detonation models are much more consistent with the distribution of the
elements inferred by modelling the spectra \citep[see, e.g.
][]{2005ApJ...623..337G}. However, the distribution of IMEs that results from the
model of 1991T is different from that produced by delayed detonation models. In
SN\,1991T there is only a small Si-dominated shell located between the interior
dominated by \Nifs\ and the external O-dominated shell (Figure
\ref{abundances1991T}).

A delayed detonation model with more complete burning leads to more $^{56}$Ni 
at the expense of O and C, also decreasing the mass of Si. For example, model 
DD\_005 from \cite{2007AA...464..683R} shows a reasonable agreement in the total 
mass of the principal elements. But looking more carefully {there
 are still some minor problems} comparing it with the abundances of our model
 of SN\,1991T (Figure~\ref{abundances1991T}). In 
{their model} the C mass is consistent with the upper limits
 from our models, but the O mass is too low and the IMEs mass is still too high.
{The behaviour of this model, where an almost complete burning
 of C implies also a large depletion of O and a large production of IMEs is a
 common behaviour of strong delayed detonation models}. This does not make it easy 
 to explain the large amount of O {that we found} in SN\,1991T
 together with the low mass of IMEs and C {in the framework 
of delayed detonation models}.
In the model TURB7 of \cite{2009ApJ...695.1257B} the elemental abundances are
very similar to those of DD\_005. Although the final abundances are not 
{too}
different from our results, the distribution of elements in TURB7 is very 
different. In the TURB7 model there is an excess of Fe-group elements in the 
outer layers. The IMEs reach a maximum in the outer layers, while in our models 
the maximum is reached at intermediate velocities, both in the delayed detonation
model and in W7. The O abundance has a maximum in an intermediate shell at an 
enclosed mass of $1\Msun$, while in our model of SN\,1991T there is no more O at 
that depth. The distribution of elements of SN\,1991T matches well the 
distribution of the TURB7 model just at the end of the detonation before most of 
the mixing when the ejecta are not yet in homologous expansion. The inversion of 
the IMEs- and the O-rich layers {in TURB7} takes place because the former has a higher 
velocity than the latter. According to \cite{2009ApJ...695.1257B} the mixing 
that changes the abundances after the end of the detonation  
happens {in any model with a detonation that} propagates inwards from a point close 
to the surface.  Indeed, the delayed detonation model ``case $c$'' 
from \cite{2005ApJ...623..337G}, which has a central detonation ignition, has 
much more stratified ejecta. The model has also the right velocity of the 
Si-rich shell. However, the explosion model ``case $c$'' has too much Si and too 
much unburned C. Moreover, \Nifs\ does not reach the outer layers as it does in  
SN\,1991T.
A slightly off-centre ignition may accomplish some more mixing, so some Fe group
elements could end up in the outer parts of the ejecta, while keeping a
well-defined Si-rich shell.
The 3D delayed detonation models from \cite{2013MNRAS.429.1156S} 
have a stratification of the outer part of the ejecta in fair agreement with 
what is inferred from {our model of} SN\,1991T. In particular, the models with few ignition 
spots show a larger production of \Nifs\ at the expenses of IMEs. The inner 
parts, however, show significant mixing of stable Fe with \Nifs. This is claimed 
to be a 3D effect of the deflagration phase that does not appear in 2D models. 
However, this does not agree with the results of the nebular spectra analysis of 
 {our model of} SN\,1991T (and other SNe\,Ia) where stable Fe is concentrated in the centre. 
In conclusion, strong delayed detonation models are good candidates for roughly
explaining the distribution of the elements {inferred from our
 model of} SN\,1991T, but the finer details
are hard to explain with the current explosion models.

\section{Conclusions}
\label{Conclusions}
 
SN\,1991T was a peculiar, overluminous supernova, well observed from early
phases. It can be considered a good representative of a class of extreme
SNe\,Ia.
Although SN\,1991T is extreme among nearby SNe\,Ia, we expect supernovae of this
subclass to be common at cosmological distances, because they are selected by
the Malmquist bias.

We obtained the abundance distribution of SN\,1991T modelling the optical
spectra in the photospheric and the nebular phase. The distribution of elements in SN\,1991T consists of an
inner layer rich of stable Fe, a large intermediate layer composed mainly of
\Nifs\ and an outer zone rich of O. IMEs are dominant only in a narrow shell at
the interface between the $^{56}$Ni-rich and O-rich zones.

Shortage of IMEs is a major peculiarity of SN\,1991T that arises from our
analysis.  Comparing the resulting abundance profile with explosion models from
the literature suggests that deflagration models do not produce enough burning
and mix up the elements too much.
We checked the consistency of
the bolometric light curve and of the kinetic energy of the assumed density
profile of the ejecta. 

Delayed detonation models are better suited for the SN\,1991T. To
produce enough \Nifs\ a delayed detonation model needs an early transition to a 
detonation. However, such models may burn O to IMEs too efficiently and at the
same time leave too much unburned C in external layers.

We see no need for a mass exceeding the Chandrasekhar limit from a combined
photospheric and nebular approach.
The modelling allows to set a convincing upper limit on the luminosity of the
 object. Our best model has a luminosity of M$_B = -19.37$.

A differential study of SN\,1991T and other similarly luminous but
spectroscopically normal SNe\,Ia (\eg SN\,1999ee) may shed more light on the
differences between them and the fine details of these extreme explosions.

\section*{Acknowledgements} 

This work has made use of the SUSPECT SN spectral archive.
This work has made use of the WISeREP data repository.
For the UV spectra we make use of the MAST archive.

We thank the Mongolian Ger Camps for the hospitality while this work was
finished in the Gobi desert.

P.M., E.C., S.B. are partially supported by the PRIN-INAF 2011 with the
 project "Transient Universe: from ESO Large to PESSTO".

K.N. acknowledges the support from the Grant-in-Aid for Scientific
Research (23224004, 23540262, 26400222) from the Japan Society for the
Promotion of Science, and the World Premier International Research
Center Initiative (WPI Initiative), MEXT, Japan.


\bibliography{biblionew}


\bsp

\label{lastpage}

\newpage
\onecolumn
\appendix
\section{Magnitudes}

Table \ref{mag_table} reports the magnitudes used for the calculation of the
bolometric light curve. The codes for the source column are as follow: 0,1 =
\cite{2004MNRAS.349.1344A}, 2 = \cite{1998AJ....115..234L}, 3
\cite{1994ApJ...434L..19S} and the IAUCs
 \cite{1991IAUC.5309....3G,1991IAUC.5273....1B,1991IAUC.5256....1W,1991IAUC.5253....2B,1991IAUC.5270....3H,1991IAUC.5246....2M}.

\begin{longtable}{c cccccc}
\label{mag_table}

   JD + 2400000   &    U     &      B     &      V     &      R     &      I       &    Source \\

\hline

%
%
%
%
%
%
48362.85&$ 12.26  \pm .02  $&$ 12.98  \pm .02  $&$ 12.84  \pm .02  $&$ 12.67  \pm .02  $&$ 12.87  \pm .02  $& 1   \\
48363.65&$       \cdots    $&$ 12.73  \pm .01  $&$ 12.61  \pm .01  $&$       \cdots    $&$       \cdots    $& 1   \\
48363.77&$       \cdots    $&$ 12.712 \pm .016 $&$ 12.618 \pm .016 $&$ 12.526 \pm .015 $&$       \cdots    $& 2   \\%
48363.80&$ 12.04  \pm .02  $&$ 12.80  \pm .02  $&$ 12.74  \pm .02  $&$       \cdots    $&$       \cdots    $& 1   \\
48364.46&$ 11.80  \pm n/a  $&$ 12.51  \pm n/a  $&$ 12.40  \pm n/a  $&$       \cdots    $&$       \cdots    $ & IAUC5246 \\
48364.64&$       \cdots    $&$ 12.46  \pm .01  $&$ 12.34  \pm .01  $&$ 12.32  \pm .01  $&$       \cdots    $& 1   \\
48364.76&$       \cdots    $&$ 12.449 \pm .016 $&$ 12.362 \pm .016 $&$ 12.299 \pm .015 $&$       \cdots    $& 2   \\%
48365.65&$       \cdots    $&$       \cdots    $&$       \cdots    $&$ 12.159 \pm .015 $&$       \cdots    $& 2   \\%
48365.66&$ 11.610 \pm .026 $&$ 12.362 \pm .016 $&$ 12.256 \pm .016 $&$ 12.134 \pm .015 $&$ 12.215 \pm .017 $& 2   \\%
48366.67&$       \cdots    $&$ 12.06  \pm .01  $&$ 12.00  \pm .01  $&$ 11.97  \pm .01  $&$       \cdots    $& 1   \\
48366.67&$       \cdots    $&$ 12.152 \pm .016 $&$ 12.061 \pm .016 $&$ 11.971 \pm .015 $&$       \cdots    $& 2   \\%
48367.65&$       \cdots    $&$       \cdots    $&$ 11.89  \pm .01  $&$       \cdots    $&$       \cdots    $& 1   \\
48367.77&$ 11.09  \pm n/a  $&$ 11.75  \pm n/a  $&$ 11.69  \pm n/a  $&$       \cdots    $&$       \cdots    $ & IAUC5270 \\
48367.77&$ 11.09  \pm n/a  $&$ 11.76  \pm n/a  $&$ 11.70  \pm n/a  $&$       \cdots    $&$       \cdots    $ & IAUC5270 \\
48368.02&$ 11.14  \pm n/a  $&$ 11.94  \pm n/a  $&$ 11.84  \pm n/a  $&$ 11.76  \pm n/a  $&$       \cdots    $ & IAUC5246 \\
48368.70&$       \cdots    $&$ 11.78  \pm .01  $&$ 11.80  \pm .01  $&$ 11.70  \pm .01  $&$       \cdots    $& 1   \\
48372.70&$       \cdots    $&$ 11.748 \pm .016 $&$       \cdots    $&$ 11.533 \pm .015 $&$ 11.693 \pm .017 $& 2   \\%
48374.64&$ 11.282 \pm .026 $&$ 11.719 \pm .016 $&$ 11.573 \pm .016 $&$ 11.511 \pm .015 $&$ 11.690 \pm .017 $& 2   \\%
48374.64&$ 11.304 \pm .026 $&$ 11.707 \pm .016 $&$ 11.560 \pm .016 $&$ 11.519 \pm .015 $&$ 11.694 \pm .017 $& 2   \\%
48375.01&$ 11.01  \pm n/a  $&$ 11.66  \pm n/a  $&$ 11.52  \pm n/a  $&$ 11.41  \pm n/a  $&$       \cdots    $ & IAUC5253 \\
48375.65&$ 11.270 \pm .026 $&$ 11.720 \pm .016 $&$ 11.537 \pm .016 $&$ 11.509 \pm .015 $&$ 11.676 \pm .017 $& 2   \\%
48375.65&$ 11.370 \pm .026 $&$ 11.715 \pm .016 $&$ 11.568 \pm .016 $&$ 11.529 \pm .015 $&$ 11.689 \pm .017 $& 2   \\%
48376.69&$ 11.233 \pm .026 $&$ 11.741 \pm .016 $&$ 11.527 \pm .016 $&$ 11.475 \pm .015 $&$       \cdots    $& 2   \\%
48376.69&$ 11.279 \pm .026 $&$ 11.696 \pm .016 $&$ 11.542 \pm .016 $&$ 11.474 \pm .015 $&$       \cdots    $& 2   \\%
48376.69&$       \cdots    $&$       \cdots    $&$       \cdots    $&$       \cdots    $&$ 11.690 \pm .017 $& 2   \\%
48376.97&$ 11.21  \pm n/a  $&$ 11.75  \pm n/a  $&$ 11.49  \pm n/a  $&$ 11.36  \pm n/a  $&$ 11.69  \pm n/a  $ & IAUC5256 \\
48377.62&$ 11.329 \pm .026 $&$ 11.694 \pm .016 $&$ 11.511 \pm .016 $&$ 11.435 \pm .015 $&$ 11.664 \pm .017 $& 2   \\%
48377.62&$       \cdots    $&$ 11.711 \pm .016 $&$ 11.482 \pm .016 $&$ 11.435 \pm .015 $&$ 11.665 \pm .017 $& 2   \\%
48380.69&$       \cdots    $&$         \cdots  $&$ 11.529 \pm .016 $&$ 11.479 \pm .015 $&$ 11.805 \pm .018 $& 2   \\%
48380.70&$       \cdots    $&$ 11.832 \pm .016 $&$ 11.525 \pm .016 $&$ 11.466 \pm .015 $&$ 11.788 \pm .017 $& 2   \\%
48381.60&$       \cdots    $&$ 11.836 \pm .016 $&$ 11.535 \pm .016 $&$ 11.518 \pm .015 $&$ 11.847 \pm .017 $& 2   \\%
48381.60&$       \cdots    $&$ 11.857 \pm .016 $&$       \cdots    $&$ 11.501 \pm .015 $&$ 11.846 \pm .017 $& 2   \\%
48382.67&$       \cdots    $&$ 11.912 \pm .016 $&$ 11.577 \pm .016 $&$ 11.565 \pm .015 $&$ 11.915 \pm .017 $& 2   \\%
48382.68&$       \cdots    $&$ 11.919 \pm .016 $&$ 11.575 \pm .016 $&$ 11.552 \pm .015 $&$ 11.917 \pm .017 $& 2   \\%
48382.77&$ 11.78  \pm n/a  $&$ 11.98  \pm n/a  $&$ 11.61  \pm n/a  $&$       \cdots    $&$       \cdots    $ & IAUC5273 \\
48383.58&$ 11.652 \pm .026 $&$ 11.959 \pm .016 $&$ 11.605 \pm .016 $&$ 11.589 \pm .015 $&$ 11.966 \pm .017 $& 2   \\%
48383.58&$       \cdots    $&$ 11.971 \pm .016 $&$ 11.583 \pm .016 $&$ 11.585 \pm .015 $&$ 11.967 \pm .017 $& 2   \\%
48385.62&$       \cdots    $&$       \cdots    $&$       \cdots    $&$ 11.729 \pm .015 $&$ 12.049 \pm .017 $& 2   \\%
48385.63&$       \cdots    $&$ 12.137 \pm .016 $&$ 11.643 \pm .016 $&$ 11.730 \pm .015 $&$ 12.051 \pm .017 $& 2   \\%
48386.55&$         \cdots  $&$ 12.208 \pm .016 $&$       \cdots    $&$       \cdots    $&$       \cdots    $& 2   \\%
48386.55&$       \cdots    $&$ 12.248 \pm .016 $&$ 11.764 \pm .016 $&$ 11.811 \pm .015 $&$ 12.087 \pm .017 $& 2   \\%
48388.57&$ 12.381 \pm .026 $&$ 12.408 \pm .016 $&$ 11.873 \pm .016 $&$ 11.917 \pm .015 $&$       \cdots    $& 2   \\%
48390.50&$ 12.750 \pm .026 $&$ 12.603 \pm .016 $&$ 11.974 \pm .016 $&$ 11.982 \pm .015 $&$ 12.076 \pm .017 $& 2   \\%
48390.50&$ 12.778 \pm .026 $&$ 12.600 \pm .016 $&$ 11.971 \pm .016 $&$ 11.976 \pm .015 $&$ 12.062 \pm .017 $& 2   \\%
48393.62&$ 13.222 \pm .026 $&$ 12.965 \pm .016 $&$ 12.150 \pm .016 $&$ 12.029 \pm .015 $&$ 12.022 \pm .017 $& 2   \\%
48395.55&$       \cdots    $&$ 13.131 \pm .016 $&$ 12.223 \pm .016 $&$       \cdots    $&$       \cdots    $& 2   \\%
48395.56&$       \cdots    $&$ 13.149 \pm .016 $&$ 12.247 \pm .016 $&$       \cdots    $&$       \cdots    $& 2   \\%
48396.56&$       \cdots    $&$ 13.39  \pm .01  $&$ 12.32  \pm .01  $&$ 12.32  \pm .01  $&$       \cdots    $& 1   \\
48396.56&$       \cdots    $&$ 13.48  \pm .01  $&$ 12.29  \pm .01  $&$       \cdots    $&$       \cdots    $& 1   \\
48396.56&$       \cdots    $&$       \cdots    $&$ 12.45  \pm .01  $&$       \cdots    $&$       \cdots    $& 1   \\
48405.59&$       \cdots    $&$ 14.071 \pm .016 $&$ 12.731 \pm .016 $&$ 12.378 \pm .015 $&$ 12.098 \pm .017 $& 2   \\%
48405.59&$       \cdots    $&$ 14.071 \pm .017 $&$ 12.735 \pm .016 $&$ 12.379 \pm .015 $&$ 12.111 \pm .017 $& 2   \\%
48414.62&$ 14.610 \pm .026 $&$ 14.302 \pm .016 $&$ 13.123 \pm .016 $&$ 12.820 \pm .015 $&$ 12.421 \pm .017 $& 2   \\%
48414.62&$       \cdots    $&$ 14.307 \pm .016 $&$       \cdots    $&$       \cdots    $&$       \cdots    $& 2   \\%
48420.47&$ 14.862 \pm .026 $&$ 14.556 \pm .016 $&$ 13.423 \pm .016 $&$ 13.102 \pm .015 $&$ 12.819 \pm .017 $& 2   \\%
48427.67&$       \cdots    $&$ 14.77  \pm .01  $&$ 13.60  \pm .01  $&$ 13.51  \pm .01  $&$       \cdots    $& 1   \\
48427.67&$       \cdots    $&$       \cdots    $&$ 13.59  \pm .01  $&$       \cdots    $&$       \cdots    $& 1   \\
48427.67&$       \cdots    $&$       \cdots    $&$ 13.66  \pm .01  $&$ 13.45  \pm .01  $&$       \cdots    $& 1   \\
48431.49&$ 15.032 \pm .030 $&$ 14.770 \pm .017 $&$ 13.744 \pm .016 $&$ 13.527 \pm .015 $&$ 13.386 \pm .017 $& 2   \\%
48431.49&$ 15.070 \pm .030 $&$ 14.746 \pm .017 $&$ 13.754 \pm .016 $&$       \cdots    $&$       \cdots    $& 2   \\%
48438.50&$       \cdots    $&$ 14.863 \pm .016 $&$ 13.936 \pm .016 $&$       \cdots    $&$ 13.712 \pm .017 $& 2   \\%
48439.47&$       \cdots    $&$ 14.919 \pm .016 $&$ 14.018 \pm .016 $&$ 13.824 \pm .015 $&$ 13.749 \pm .017 $& 2   \\%
48439.47&$       \cdots    $&$ 14.931 \pm .016 $&$ 13.986 \pm .016 $&$ 13.830 \pm .015 $&$ 13.738 \pm .017 $& 2   \\%
48439.83&$       \cdots    $&$ 14.91  \pm n/a  $&$ 14.03  \pm n/a  $&$       \cdots    $&$       \cdots    $ & IAUC5309 \\
48440.46&$       \cdots    $&$ 14.897 \pm .016 $&$ 14.059 \pm .016 $&$ 13.859 \pm .001 $&$ 13.821 \pm .017 $& 2   \\%
48440.46&$       \cdots    $&$ 14.917 \pm .016 $&$ 14.058 \pm .016 $&$ 13.872 \pm .015 $&$ 13.790 \pm .017 $& 2   \\%
48440.87&$       \cdots    $&$ 14.92  \pm n/a  $&$ 14.00  \pm n/a  $&$       \cdots    $&$       \cdots    $ & IAUC5309 \\
48442.48&$ 15.265 \pm .026 $&$ 14.888 \pm .016 $&$ 14.045 \pm .016 $&$ 13.888 \pm .015 $&$ 13.778 \pm .017 $& 2   \\%
48445.46&$       \cdots    $&$       \cdots    $&$ 14.139 \pm .016 $&$       \cdots    $&$       \cdots    $& 2   \\%
48445.47&$       \cdots    $&$ 14.934 \pm .017 $&$ 14.143 \pm .016 $&$       \cdots    $&$       \cdots    $& 2   \\%
48459.49&$       \cdots    $&$ 15.115 \pm .016 $&$ 14.476 \pm .016 $&$ 14.407 \pm .015 $&$ 14.538 \pm .017 $& 2   \\%
48469.51&$       \cdots    $&$ 15.42  \pm .02  $&$       \cdots    $&$ 14.78  \pm .02  $&$       \cdots    $& 1   \\
48480.49&$       \cdots    $&$       \cdots    $&$ 15.29  \pm .02  $&$       \cdots    $&$       \cdots    $& 1   \\
48593.46&$       \cdots    $&$ 16.95  \pm .06  $&$ 17.01  \pm .05  $&$ 17.59  \pm .06  $&$       \cdots    $& 0   \\
48593.46&$       \cdots    $&$ 16.98  \pm .06  $&$ 17.01  \pm .05  $&$ 17.62  \pm .06  $&$       \cdots    $& 0   \\
48594.46&$       \cdots    $&$ 17.06  \pm .06  $&$ 17.10  \pm .06  $&$ 17.65  \pm .06  $&$       \cdots    $& 0   \\
48597.8 &$       \cdots    $&$ 17.20  \pm .03  $&$  17.13 \pm .03  $&$  17.65 \pm .05  $&$  17.34 \pm .08  $& 3   \\%
48603.46&$       \cdots    $&$ 17.04  \pm .06  $&$ 17.07  \pm .06  $&$ 17.68  \pm .06  $&$       \cdots    $& 0   \\
48628.83&$       \cdots    $&$ 17.496 \pm .019 $&$ 17.484 \pm .018 $&$       \cdots    $&$ 17.939 \pm .040 $& 2   \\%
48629.83&$       \cdots    $&$ 17.513 \pm .018 $&$ 17.475 \pm .018 $&$       \cdots    $&$ 17.951 \pm .046 $& 2   \\%
48653.90&$       \cdots    $&$ 17.66  \pm .05  $&$ 17.95  \pm .05  $&$ 18.54  \pm .05  $&$       \cdots    $& 1   \\
48657.60&$       \cdots    $&$       \cdots    $&$ 17.88  \pm .07  $&$       \cdots    $&$       \cdots    $& 0   \\
48657.82&$       \cdots    $&$ 17.74  \pm .04  $&$ 17.96  \pm .03  $&$ 18.56  \pm .04  $&$       \cdots    $& 1   \\
48658.66&$       \cdots    $&$ 17.75  \pm .04  $&$ 17.94  \pm .08  $&$ 18.52  \pm .06  $&$       \cdots    $& 0   \\
48677.78&$       \cdots    $&$       \cdots    $&$ 18.116 \pm .022 $&$       \cdots    $&$ 18.367 \pm .045 $& 2   \\%
48681.78&$       \cdots    $&$ 18.02  \pm .06  $&$ 18.34  \pm .05  $&$ 18.97  \pm .05  $&$       \cdots    $& 1   \\
48686.46&$       \cdots    $&$ 18.25  \pm .10  $&$ 18.27  \pm .06  $&$ 19.00  \pm .08  $&$       \cdots    $& 0   \\
48687.85&$       \cdots    $&$       \cdots    $&$ 18.343 \pm .017 $&$ 19.086 \pm .022 $&$ 18.489 \pm .026 $& 2   \\%
48688.46&$       \cdots    $&$ 18.25  \pm .10  $&$ 18.35  \pm .06  $&$ 19.05  \pm .08  $&$       \cdots    $& 0   \\
48688.8 &$       \cdots    $&$       \cdots    $&$  18.35 \pm .03  $&$  19.05 \pm .05  $&$  18.43 \pm .05  $& 3   \\%
48691.78&$       \cdots    $&$ 18.445 \pm .018 $&$ 18.357 \pm .019 $&$ 19.080 \pm .025 $&$ 18.489 \pm .031 $& 2   \\%
48692.67&$       \cdots    $&$ 18.42  \pm .07  $&$ 18.42  \pm .07  $&$ 19.07  \pm .06  $&$       \cdots    $& 1   \\
48720.77&$       \cdots    $&$ 18.834 \pm .023 $&$ 18.765 \pm .022 $&$       \cdots    $&$ 18.952 \pm .034 $& 2   \\%
48731.62&$       \cdots    $&$       \cdots    $&$ 19.010 \pm .052 $&$       \cdots    $&$       \cdots    $& 2   \\%
48731.63&$       \cdots    $&$ 18.916 \pm .078 $&$ 18.947 \pm .047 $&$       \cdots    $&$       \cdots    $& 2   \\%
48742.8 &$       \cdots    $&$ 19.00  \pm .03  $&$  19.05 \pm .03  $&$  19.64 \pm .08  $&$  18.81 \pm .09  $& 3   \\%
48746.69&$       \cdots    $&$ 18.99  \pm .05  $&$ 19.12  \pm .05  $&$ 19.58  \pm .05  $&$ 19.00  \pm .08  $& 1   \\
48776.53&$ 20.346 \pm .250 $&$ 19.654 \pm .078 $&$ 19.662 \pm .057 $&$ 20.271 \pm .102 $&$ 19.374 \pm .119 $& 2   \\%
48776.54&$       \cdots    $&$ 19.773 \pm .097 $&$ 19.684 \pm .063 $&$ 20.162 \pm .097 $&$ 19.359 \pm .113 $& 2   \\%
48776.8 &$       \cdots    $&$ 19.43  \pm .03  $&$  19.49 \pm .03  $&$  20.03 \pm .08  $&$  19.08 \pm .12  $& 3   \\%
48829.67&$       \cdots    $&$ 19.97  \pm .10  $&$ 20.27  \pm .05  $&$ 20.57  \pm .05  $&$       \cdots    $& 1   \\

\end{longtable}
\twocolumn
\end{document}